\documentclass[aps,nofootinbib,floatfix,amsmath,amssymb,superscriptaddress,preprint]{revtex4-1}
\pdfoutput=1

\usepackage{preamble}

\usepackage{mathpazo}
\usepackage{tgpagella}
\usepackage[scaled=0.85 ]{beramono}

\usepackage{slashed}

\usepackage{color}
\definecolor{keyword}{HTML}{008000}
\definecolor{emph}{HTML}{0000FF}
\definecolor{string}{HTML}{A52A2A}
\definecolor{comment}{HTML}{004461}
\definecolor{back}{HTML}{F8F8F8}
\definecolor{arrow}{HTML}{745334}

\usepackage{listings}

\newcommand{\code}[1]{\begingroup\tt#1\endgroup}
\xnewcommand{\LHCOreader}{\code{LHCO\_reader}}
\xnewcommand{\LHCO}{\code{LHCO}}
\xnewcommand{\ROOT}{\code{ROOT}}  
\newcommand{\command}{\color{arrow}{\ensuremath{\bm{>\!\!>\!\!>}}}} 

\usepackage{attachfile}
\usepackage{accsupp}
\usepackage{verbatim}
\usepackage{color}
\usepackage[misc]{ifsym}
\usepackage{fancyvrb}
\makeatletter
\lst@RequireAspects{writefile}

\newenvironment{rawcode}
  {\VerbatimOut{code.py}}
  {\endVerbatimOut}

\lstnewenvironment{cliprawcode}{}
{%
  \marginpar{\attachfile[appearance=false,icon=Paperclip,mimetype=text/py]{code.py}}%
}

\lstnewenvironment{clipcode}{%
  \lst@BeginAlsoWriteFile{code.py}%
}
{%
  \lst@EndWriteFile
  \marginpar{\attachfile[appearance=false,icon=Paperclip,mimetype=text/py]{code.py}}
}

\def\add@lstline
{\xdef\lstfile{\unexpanded\expandafter{\lstfile}\the\verbatim@line\string^^J}}
\makeatother

\begin{document}

\title{\LHCOreader: A new code for reading and analyzing detector-level events stored in \LHCO format}
\date{\today}
\author{Andrew Fowlie}
\email{Andrew.Fowlie@Monash.edu.au}
\affiliation{ARC Centre of Excellence for Particle Physics at the Tera-scale, School of Physics and Astronomy, Monash University, Melbourne, Victoria 3800 Australia}

\date{\today}


\begin{abstract}
We present a new Python module --- \LHCOreader --- for reading detector-level events in \LHCO format generated from detector-simulators such as \code{PGS} and \code{Delphes}. Emphasis is placed upon ease of use. The module is installed via \code{pip}. Once installed, one can read and scrutinize events, with helpful functions and classes for cutting, plotting and inspecting events, and manipulating four-momenta by \eg boosting. Implementing complicated analyses should be straight-forward by coding any cuts into Python. Furthermore, common kinematic variables, including $\alpha_T$ and razor are included out of the box, and $m_{T2}$ and $m_{T2}^W$ are provided by interfaces with external libraries.

\end{abstract}


\maketitle

\section{Introduction}
\label{index:introduction}\label{index:lhco-reader}
In order to make use of the data from the Large Hadron Collider (LHC), we must compare it with predictions from the Standard Model\cite{Glashow:1961tr,salam:sm,Weinberg:1967tq} and models of new physics \see{Plehn:2009nd}. Only by doing so can we investigate the power of LHC experiments to reject or discover models of new physics, and infer the parameters of the Standard Model. Making such predictions is challenging, and is typically broken down into four stages:
\begin{enumerate}
\item Simulating a hard scattering at the matrix-element level at a fixed-order in perturbation theory, with, for example, 
\code{ALPGEN}\cite{Mangano:2002ea},
\code{MadGraph}\cite{Alwall:2014hca}, 
\code{Helac}\cite{Cafarella:2007pc},
\code{Whizard}\cite{Kilian:2007gr}, 
\code{JHUGen}\cite{Bolognesi:2012mm,Gao:2010qx,Gao:2010qx,JHU},
\code{Comix}\cite{Gleisberg:2008fv},
\code{ISAJET}\cite{Paige:2003mg},
\code{CompHEP}\cite{Pukhov:1999gg} or
\code{SEHRPA}\cite{Gleisberg:2008ta}.

\item ``Matching'' initial- and final-state parton showers with the matrix element, and hadronizing the resulting colored objects. This is commonly achieved with 
\code{Pythia}\cite{Sjostrand:2007gs}, 
\code{Herwig}\cite{Bahr:2008pv} or
\code{Ariadne}\cite{Lonnblad:1992tz}.

\item Detector-simulation by publicly available programs including \code{Delphes}\cite{deFavereau:2013fsa} and \code{PGS}\cite{PGS}.  A common format for detector-level events is the LHC Olympics format (\LHCO)\cite{LHCO}. \LHCO is a human-readable format supported by \code{Delphes} and \code{PGS}.

\item Finally, selections on the phase-space of the final states observed in the (simulated) detector. With judicious selections, we wish to find regions of phase-space in which we can discriminate between rival hypotheses. We refer to this as an analysis. This requires that we sift through thousands of complicated events, though is by far the simplest stage.
\end{enumerate}

\LHCOreader is a Python module for reading \LHCO files from detector simulators such as \code{PGS} and \code{Delphes} into a Python class, with useful functions for implementing an analysis. It can also read \ROOT\cite{Antcheva:2009zz} files from \code{Delphes}, by immediately converting them to \LHCO files. Having read an \LHCO file, there are many functions for implementing an analysis, including access to four-momenta, kinematic variables such as $H_T$, $\slashed{H}_T$, $\alpha_T$, razor, $m_{T2}$ and $m_{T2}^W$, plotting, sorting and cutting events, counting the numbers of types of object in an event, and calculating the angular separation between objects. Similar functionality is available in \code{MadAnalysis}\cite{Conte:2012fm,Conte:2013mea,Conte:2014xya}, \code{Seer}\cite{Martin:2015hra} and \code{CutLHCO}\cite{Walker:2012vf}, though we place emphasis on ease of use, an intuitive, object-oriented format for events and the possibility of writing an analysis in Python. \code{MadAnalysis} includes a Pythonic interactive shell, but it is restrictive: you must select predefined cuts and user-defined cuts must be written in \code{C++} in so-called expert-mode\cite{Conte:2014zja}. Python is particularly well suited for an easy-to-use program: it is a high-level, object-oriented language that is well known in the physics community and well documented online. This choice is, however, to the detriment of performance, especially compared with the \code{C++} backend in \code{MadAnalysis}. As we shall illustrate, however, it is adequate for a reasonable number of events, $\lesssim 100\,\text{k}$. 

This manual is structured as follows: we describe the \LHCO format in \refsec{Sec:LHCO}, installation of \LHCOreader in \refsec{Sec:Installation}, simple and advanced usage of \LHCOreader in \refsec{Sec:Quickstart}, present a complete analysis in \refsec{Sec:Example}, and discuss performance in \refsec{Sec:Performance} before summarizing in \refsec{Sec:Summary}. For a full description, including a description of all classes and functions, see:
\begin{lstlisting}
http://lhco-reader.readthedocs.org
\end{lstlisting}
Note that complete self-contained snippets of code can be opened in a text-editor by double-clicking the paperclip symbol --- \noattachfile[appearance=false,icon=Paperclip,mimetype=text/tex] --- in the margin of code-snippets.\footnote{This might not work in all PDF viewers, though it should work with \eg Adobe Acroread.}

\section{\LHCO format}\label{Sec:LHCO}
The \LHCO format originates from the LHC Olympics, a challenge to reconstruct physics signals from a ``black-box'' of detector-level events, organized in anticipation of the beginning of the LHC.  For every event, \LHCO includes information about the trigger, every object in the final-state and the missing transverse energy (MET). The format is a human-readable array with eleven columns, separated by at least one space or tab. The eleven columns are:
\begin{enumerate}
\item \code{\#} (integer) --- Numbers the objects in an event, beginning at zero. The zeroth object --- \code{0} --- contains only two further columns: the event number (beginning at \code{1}) and trigger information. Thus, a new event begins with \code{0} in the first column. All further objects begin with numbers greater than \code{0} and contain ten further columns, as follows.
\item \code{typ} (integer) --- Indicates the type of detector-level object:
\begin{itemize}
\item \code{0} --- Photon
\item \code{1} --- Electron
\item \code{2} --- Muon
\item \code{3} --- Hadronically-decaying tau
\item \code{4} --- Jet (defined by a jet-clustering algorithm in the detector-simulation)
\item \code{6} --- Missing transverse energy (MET). All events contain exactly one MET object.
\end{itemize}
Note that \code{5} is absent. The remaining columns define properties of the object of type--\code{typ}.
\item \code{eta} (float) --- The pseudo-rapidity, $\eta\equiv -\ln(\tan\theta/2)$, where $\theta$ is the polar angle. 
\item \code{phi} (float)  --- The azimuthal angle, $\phi$, measured around the beam-line.       
\item \code{pt} (float)  --- The transverse momentum, $p_T \equiv \sqrt{p_x^2 + p_y^2}$.
\item \code{jmass} (float)  --- Invariant mass; for a jet, this is the invariant mass of all subjet constituents.
\item \code{ntrk} (signed integer) --- Number of tracks associated with object. If the object is a lepton, the sign is the charge of the lepton.  This should always be positive for objects other than leptons.
\item \code{btag} (integer) --- Information about $b$-tags. If \code{0}, the object was not $b$-tagged.  If greater than zero, the object was $b$-tagged. The precise definitions of values greater than zero depend upon detector simulation; this is a means for recording more detailed information about $b$-tagging. This should always be \code{0} for objects other than jets.
\item \code{had/em} (float)  --- The ratio of hadronic versus electromagnetic energy deposited in the calorimeter cells. 
\item \code{dummy} --- Always \code{0}, though potentially could hold further information.
\item \code{dummy} --- Always \code{0}, though potentially could hold further information.
\end{enumerate}
Lines beginning with \code{\#} are comments. Comments and blank lines are ignored.

This is an example of a single \LHCO event:
\begin{lstlisting}
# Number Trigger
0 10000  3631
# typ  eta   phi   pt    jmass  ntrk btag had/em dummy dummy
1 1   -0.139 4.399 58.46 0.00   1.0  0.0  0.00   0.0   0.0
2 1    0.867 2.317 34.76 0.00  -1.0  0.0  0.01   0.0   0.0
3 3    0.228 6.065 19.35 0.00   3.0  0.0  1.49   0.0   0.0
4 4    2.312 0.841 30.03 5.86   7.0  0.0  1.88   0.0   0.0
5 4    2.522 1.483 22.23 3.41   8.0  0.0  1.30   0.0   0.0
6 6    0.000 3.064 4.73  0.00   0.0  0.0  0.00   0.0   0.0
\end{lstlisting}
This is the 10000-th event in a file and contains six objects: two electrons (\code{1} and \code{2}), a tau (\code{3}), two jets (\code{4} and \code{5}) and MET (\code{6}). We see that neither jet was $b$-tagged, and that the first electron was positively charged. The comments beginning with \code{\#} are ignored in the \LHCO format, but included in my example to improve legibility.

\section{Installation}\label{Sec:Installation}
This describes installation of \LHCOreader. The easiest method of installing \LHCOreader is via the \code{pip} package manager:
\begin{lstlisting}
pip install LHCO_reader
\end{lstlisting} 
on Linux or Mac OSX. This requires a $50\,\text{KB}$ download. It might, of course, be necessary to prepend that command with \code{sudo}. If you don't have \code{pip}, install it \eg via \code{aptitude} or via \code{yum}:
\begin{lstlisting}
apt-get install python-pip
yum install python-pip
\end{lstlisting}
\LHCOreader requires some common modules that you might need to install separately, the most obscure of which is \code{prettytable}\cite{pt}. To install \code{prettytable} via \code{pip}:
\begin{lstlisting}
pip install prettytable
\end{lstlisting}
\LHCOreader requires version \code{0.6} of \code{prettytable} or later. You might also need to install \code{numpy}\cite{numpy}, \code{scipy}\cite{scipy,scipy2} and \code{matplotlib}\cite{matplotlib}\footnote{\LHCOreader is functional without \code{matplotlib}, though the \code{plot()} function will result in an \code{ImportError}.} \eg via \code{aptitude} or via \code{yum}:
\begin{lstlisting}
apt-get install python-numpy python-matplotlib python-scipy
yum install python-numpy python-matplotlib python-scipy
\end{lstlisting}
\LHCOreader depends upon several standard Python modules: \code{os}, \code{sys}, \code{warnings}, \code{math}, \code{collections}, \code{subprocess} and \code{inspect}. 

\subsection{Advanced installation}
If installation via \code{pip} is problematic \eg for Windows, clone the source code from \code{git-hub}:
\begin{lstlisting}
git clone https://github.com/innisfree/LHCO_reader.git
\end{lstlisting}
or download the source code via a web-browser from 
\begin{lstlisting}
https://github.com/innisfree/LHCO_reader/archive/master.zip
\end{lstlisting}
The module requires no installation, though in this case, you must, of course, be in the correct directory to import the module.

\subsection{Reporting bugs}
Please report all bugs, counter-intuitive behavior or feature requests, preferably via \code{git-hub}
\begin{lstlisting}
https://github.com/innisfree/LHCO_reader/issues
\end{lstlisting}
but alternatively via e-mail to \url{Andrew.Fowlie@Monash.Edu.Au}.

\section{Quick-start}\label{Sec:Quickstart}
\subsection{Importing module and loading events}
Having obtained the package, download the \code{example.lhco} set of detector-level events in a new directory:
\begin{clipcode}
mkdir test_LHCO_reader
cd test_LHCO_reader
wget https://raw.githubusercontent.com/innisfree/LHCO_reader/master/example.lhco
\end{clipcode}
The module is written in Python-2 and tested with Python-2.7.6. Within a \code{python} shell in the same directory as your events, execute the commands:\footnote{
If you install via \code{pip}, this is possible from any directory, whereas if you download the source
code, make sure you a single directory above the \LHCOreader directory.}
\begin{rawcode}
from LHCO_reader import LHCO_reader
events = LHCO_reader.Events(f_name="example.lhco")
print events
\end{rawcode}
\begin{cliprawcode}
&\command& from LHCO_reader import LHCO_reader
&\command& events = LHCO_reader.Events(f_name="example.lhco")
&\command& print events
+------------------+--------------+
| Number of events | 10000        |
| Description      | example.lhco |
+------------------+--------------+
\end{cliprawcode}
The first line imports the module. The second line loads the events in \code{example.lhco} into an \code{Events()} class (you should replace \code{example.lhco} with its full path if that file is not inside your present working directory). The third line prints information about the events --- we find that there are 10000 events in the file with description \code{example.lhco}. The keyword \code{f\_name} code be a \ROOT or \LHCO file. There are other optional arguments, including \code{n\_events}, which reads a maximum number of events from an \LHCO file.

\subsection{Structure of \code{Events()} class}
Suppose we begin by loading an \LHCO file into an \code{Events()} class, 
\begin{rawcode}
from LHCO_reader import LHCO_reader
events = LHCO_reader.Events(f_name="example.lhco")
\end{rawcode}
\begin{cliprawcode}
&\command& from LHCO_reader import LHCO_reader
&\command& events = LHCO_reader.Events(f_name="example.lhco")
\end{cliprawcode}
An instance of an \code{Events()} class, in this case \code{events}, is structured as follows:
\begin{itemize}
\item \code{events}  --- A list-like \code{Events()} class of all events in the \LHCO file, \eg
\begin{rawcode}
from LHCO_reader import LHCO_reader
events = LHCO_reader.Events(f_name="example.lhco")
print events
\end{rawcode}
\begin{cliprawcode}
&\command& print events
+------------------+--------------+
| Number of events | 10000        |
| Description      | example.lhco |
+------------------+--------------+
\end{cliprawcode}
\item \code{events[0]} --- The zeroth event in the \LHCO file contains many objects:
\begin{rawcode}
from LHCO_reader import LHCO_reader
events = LHCO_reader.Events(f_name="example.lhco")
print events[0]
\end{rawcode}
\begin{cliprawcode}
&\command& print events[0]
+----------+--------+-------+--------+-------+------+------+-------+
|  Object  |  eta   |  phi  |   PT   | jmass | ntrk | btag | hadem |
+----------+--------+-------+--------+-------+------+------+-------+
| electron | -0.745 | 4.253 | 286.72 |  0.0  | -1.0 | 0.0  |  0.0  |
| electron | -0.073 | 4.681 | 44.56  |  0.0  | 1.0  | 0.0  |  0.0  |
|   jet    | -0.565 | 1.126 | 157.44 | 12.54 | 16.0 | 0.0  |  0.57 |
|   jet    | -0.19  | 1.328 | 130.96 |  12.3 | 18.0 | 0.0  | 10.67 |
|   jet    | 0.811  | 6.028 | 17.49  |  3.47 | 8.0  | 0.0  |  2.37 |
|   jet    | 0.596  | 0.853 | 12.47  |  2.53 | 7.0  | 0.0  |  1.26 |
|   jet    | -1.816 | 0.032 |  6.11  |  1.18 | 0.0  | 0.0  |  0.56 |
|   jet    | 0.508  | 1.421 |  6.01  |  0.94 | 7.0  | 0.0  |  2.59 |
|   MET    |  0.0   | 2.695 | 21.43  |  0.0  | 0.0  | 0.0  |  0.0  |
+----------+--------+-------+--------+-------+------+------+-------+
\end{cliprawcode}
The events can be looped with e.g.:
\begin{rawcode}
from LHCO_reader import LHCO_reader
events = LHCO_reader.Events(f_name="example.lhco")
for event in events:
    print event
\end{rawcode}
\begin{cliprawcode}
&\command& for event in events:
&\command&     ... scrutinize an event ...
\end{cliprawcode}
but beware that altering list-type objects in a loop can be problematic. The best way to cut events is with the \code{cut()} function, described shortly. The events themselves, \eg \code{events[0]}, are \code{Event()} classes.
    
\item\code{events[0]["electron"]} --- A list of all electrons in the zeroth event in the \LHCO file:
\begin{rawcode}
from LHCO_reader import LHCO_reader
events = LHCO_reader.Events(f_name="example.lhco")
print events[0]["electron"]
\end{rawcode}
\begin{cliprawcode}
&\command& print events[0]["electron"]
+----------+--------+-------+--------+-------+------+------+-------+
|  Object  |  eta   |  phi  |   PT   | jmass | ntrk | btag | hadem |
+----------+--------+-------+--------+-------+------+------+-------+
| electron | -0.745 | 4.253 | 286.72 |  0.0  | -1.0 | 0.0  |  0.0  |
| electron | -0.073 | 4.681 | 44.56  |  0.0  | 1.0  | 0.0  |  0.0  |
+----------+--------+-------+--------+-------+------+------+-------+
\end{cliprawcode}
The objects themselves, \eg \code{events[0]["electron"]}, are an \code{Objects()} class --- a class for a list of objects of the same type in an event. The possible keys are \code{electron}, \code{muon}, \code{tau}, \code{jet}, \code{MET} and \code{photon}:

\item\code{events[0]["electron"][0]} --- The zeroth electron in the zeroth event in the \LHCO file:
\begin{rawcode}
from LHCO_reader import LHCO_reader
events = LHCO_reader.Events(f_name="example.lhco")
print events[0]["electron"][0]
\end{rawcode}
\begin{cliprawcode}
&\command& print events[0]["electron"][0]
+----------+--------+-------+--------+-------+------+------+-------+
|  Object  |  eta   |  phi  |   PT   | jmass | ntrk | btag | hadem |
+----------+--------+-------+--------+-------+------+------+-------+
| electron | -0.745 | 4.253 | 286.72 |  0.0  | -1.0 | 0.0  |  0.0  |
+----------+--------+-------+--------+-------+------+------+-------+
\end{cliprawcode}
The individual objects themselves, \eg \code{events[0]["electron"][0]}, are \code{Object()} classes --- a class for a single object.
  
\item\code{events[0]["electron"][0]["PT"]} --- The transverse momentum of the zeroth electron in the zeroth event in the \LHCO file:
\begin{rawcode}
from LHCO_reader import LHCO_reader
events = LHCO_reader.Events(f_name="example.lhco")
print events[0]["electron"][0]["PT"]
\end{rawcode}
\begin{cliprawcode}
&\command& print events[0]["electron"][0]["PT"]
286.72
\end{cliprawcode}
The other possible keys are \code{event}, \code{type}, \code{eta}, \code{phi}, \code{PT}, \code{jmass}, \code{ntrk}, \code{btag} and \code{hadem}. The keys correspond the column headings mentioned in the previous section. The properties, \eg \code{events[0]["electron"][0]["PT"]}, are simply floats.
\end{itemize}
All dimensional quantities in \LHCOreader are in \gev.

\subsection{Cutting events}
A cut removes events that fail a particular criterion. Cuts can be implemented with one-line lambda-functions or ordinary functions. The function describing a cut should return \code{True} if the event should be cut, and \code{False} otherwise. As an example, to cut events with one tau-lepton:\footnote{We show the output from running the code snippet in a script. The format of the output differs slightly in an interactive shell.}
\begin{rawcode}
from LHCO_reader import LHCO_reader
events = LHCO_reader.Events(f_name="example.lhco")
tau = lambda event: event.number()["tau"] == 1
events.cut(tau)
print events
\end{rawcode}
\begin{cliprawcode}
&\command& tau = lambda event: event.number()["tau"] == 1
&\command& events.cut(tau)
&\command& print events
+------------------+--------------+
| Number of events | 8656         |
| Description      | example.lhco |
+------------------+--------------+

+------------------------------------------------+------------------+
| tau = lambda event: event.number()["tau"] == 1 | 0.8656           |
| Combined acceptance                            | 0.8656           |
| 68% interval                                   | [0.8621, 0.8690] |
+------------------------------------------------+------------------+
\end{cliprawcode}
The first line defines a function, \code{tau()}, that returns \code{True} if and only if there is exactly one tau-lepton in an event. The second line applies that cut to the events, removing events with one tau-lepton. We find that $0.856$ of events remain. 

The efficiency is estimated from binomial statistics as we are estimating the probability of a success, $p$, given $k$ successes from $n$ trials. The efficiency, $\epsilon$, plays the role of the probability of a success, the number of events plays the role of the number of trials and the number of accepted events plays the role of the number of successes. We estimate the efficiency, $\epsilon$, as
\begin{equation}
\hat\epsilon \equiv \frac{k}{n}
\end{equation}
The estimator $\hat\epsilon$ is the maximum likelihood estimator and is such that $\langle \hat \epsilon\rangle = \epsilon$.
A common approach is to estimate the error as
\begin{equation}\label{Eq:Crude}
\Delta \hat\epsilon = \sqrt{\frac{\hat\epsilon (1 - \hat\epsilon)}{n}}.
\end{equation}
from a binomial variance \see{Conte:2012fm,Conte:2013mea,Conte:2014xya}. This, 
however, performs poorly and under-covers unless $n\epsilon$ is large \see{agresti2003categorical}. 
Following Agresti\cite{agresti2003categorical}, 
by default we utilize the Clopper-Pearson exact binomial test\cite{CLOPPER01121934} to 
construct a $1\sigma$ confidence interval for the efficiency, which in fact
slightly over-covers, \ie it is conservative. This is an attribute of an 
\code{Events()} class, \eg \code{events.interval\_acceptance()}. This attribute has an optional keyword,
\code{crude}, \code{False} by default, which switches between the crude
binomial approach in \refeq{Eq:Crude} and Clopper-Pearson. 
In the limit of large $n\epsilon$, the
difference is small as \refeq{Eq:Crude} achieves correct coverage by the central-limit theorem.

Every event includes a function \code{number()}, that returns a dictionary of the numbers of objects in that event:
\begin{rawcode}
from LHCO_reader import LHCO_reader
events = LHCO_reader.Events(f_name="example.lhco")
print events[0].number()
print events[0].number()["electron"]
\end{rawcode}
\begin{cliprawcode}
&\command& print events[0].number()
+--------+----------+------+-----+-----+-----+
| photon | electron | muon | tau | jet | MET |
+--------+----------+------+-----+-----+-----+
|   0    |    2     |  0   |  0  |  6  |  1  |
+--------+----------+------+-----+-----+-----+
&\command& print events[0].number()["electron"]
2
\end{cliprawcode}
Individual numbers of events can be readily obtained, as in the final line. The total number of objects in an event is found with the \code{multiplicity()} function. The \code{number()} functions include an optional keyword \code{anti\_lepton}, \code{False} by default, indicating that leptons and anti-leptons should be counted separately. 

You could, of course, define complicated cut-functions based upon all manner of properties, with ordinary functions such as 
\begin{rawcode}
from LHCO_reader import LHCO_reader
events = LHCO_reader.Events(f_name="example.lhco")

def mycut(event):
    # Check whether any jets
    if event.number()["jet"] == 0:
        return False 
    # Check hardest jet  
    event["jet"].order("PT")
    return event["jet"][0] < 50.
     
events.cut(mycut)
\end{rawcode}
\begin{cliprawcode}
&\command& def mycut(event):
&\command&     # Check whether any jets
&\command&     if event.number()["jet"] == 0:
&\command&         return False 
&\command&     # Check hardest jet  
&\command&     event["jet"].order("PT")
&\command&     return event["jet"][0] < 50.
&\command&     
&\command& events.cut(mycut)
\end{cliprawcode}
This function cuts events with no jets and events in which the hardest jet is less than $50\gev$. 

\subsection{Leptons and anti-leptons, and $b$-jets}
In general, leptons and anti-leptons are not distinguished by the structure of the \code{Events()} class. This is often convenient --- common $P_T$ and $\eta$ cuts apply to leptons of either charge. There are certainly situations, however, in which they ought to be distinguished. As previously mentioned, the \code{number()} functions include an optional keyword \code{anti\_lepton}, which, if \code{True}, distinguishes lepton charges:
\begin{rawcode}
from LHCO_reader import LHCO_reader
events = LHCO_reader.Events(f_name="example.lhco")

print(events[0].number())
print(events[0].number(anti_lepton=True))
\end{rawcode}
\begin{cliprawcode}
&\command& print(events[0].number())
+--------+----------+------+-----+-----+-----+
| photon | electron | muon | tau | jet | MET |
+--------+----------+------+-----+-----+-----+
|   0    |    2     |  0   |  0  |  6  |  1  |
+--------+----------+------+-----+-----+-----+
&\command& print(events[0].number(anti_lepton=True))
+--------+----------+------+-----+-----+-----+---------------+-----------+----------+
| photon | electron | muon | tau | jet | MET | anti-electron | anti-muon | anti-tau |
+--------+----------+------+-----+-----+-----+---------------+-----------+----------+
|   0    |    1     |  0   |  0  |  6  |  1  |       1       |     0     |    0     |
+--------+----------+------+-----+-----+-----+---------------+-----------+----------+
\end{cliprawcode}
This function returns a dictionary --- \eg you could access the number of anti-electrons in the zeroth event with \code{events[0].number(anti\_lepton=True)["anti-electron"]}.

Furthermore, each \code{Objects()} class, \eg \code{events[0]["electron"]}, contains a \code{pick\_charge()} function, with a mandatory argument --- the required charge. This function returns another \code{Objects()} class containing only objects of specified charge. For example,
\begin{rawcode}
from LHCO_reader import LHCO_reader
events = LHCO_reader.Events(f_name="example.lhco")

# Look at all electrons
print(events[0]["electron"])
# Only look at anti-electrons with charge 1
print(events[0]["electron"].pick_charge(1))
\end{rawcode}
\begin{cliprawcode}
&\command& # Look at all electrons
&\command& print(events[0]["electron"])
+----------+--------+-------+--------+-------+------+------+-------+
|  Object  |  eta   |  phi  |   PT   | jmass | ntrk | btag | hadem |
+----------+--------+-------+--------+-------+------+------+-------+
| electron | -0.745 | 4.253 | 286.72 |  0.0  | -1.0 | 0.0  |  0.0  |
| electron | -0.073 | 4.681 | 44.56  |  0.0  | 1.0  | 0.0  |  0.0  |
+----------+--------+-------+--------+-------+------+------+-------+
&\command& # Only look at anti-electrons with charge 1
&\command& print(events[0]["electron"].pick_charge(1))
+----------+--------+-------+-------+-------+------+------+-------+
|  Object  |  eta   |  phi  |   PT  | jmass | ntrk | btag | hadem |
+----------+--------+-------+-------+-------+------+------+-------+
| electron | -0.073 | 4.681 | 44.56 |  0.0  | 1.0  | 0.0  |  0.0  |
+----------+--------+-------+-------+-------+------+------+-------+
\end{cliprawcode}

This function utilizes a \code{charge()} function, belonging to an \code{Object()} class. This function, \eg \code{event[0]["jet"][0].charge()}, returns the charge of an object. If the object is not a lepton, it returns \code{None}.

In a manner similar to that in the \code{pick\_charge()} function, it is possible to build an \code{Objects()} class of $b$-tagged jets with the function \code{pick\_b\_jets()}, which belongs to an \code{Event()} class, \eg \code{event[0].pick\_b\_jets()} returns an  \code{Objects()} class of $b$-tagged jets. This function includes an optional keyword \code{tagged}, \code{True} by default, with which one can find ordinary, non-$b$-tagged jets: \code{pick\_b\_jets(tagged=False)}. Furthermore, the function \code{count\_b\_jets()} returns the number of $b$-tagged jets in an event, \eg \code{event[0].count\_b\_jets()}.

\subsection{Using four-momenta}
When implementing an analysis, one might wish to use four-momenta, and consider \eg Minkowski-products. This is possible, as every event contains a \code{vector()} function, which returns a \code{Fourvector()} class:
\begin{rawcode}
from LHCO_reader import LHCO_reader
events = LHCO_reader.Events(f_name="example.lhco")

# Look at four-momentum of two jets in zeroth event
j0 = events[0]["jet"][0].vector()
j1 = events[0]["jet"][1].vector()
\end{rawcode}
\begin{cliprawcode}
&\command& # Look at four-momentum of two jets in zeroth event
&\command& j0 = events[0]["jet"][0].vector()
&\command& j1 = events[0]["jet"][1].vector()
\end{cliprawcode}
The objects \code{j0} and \code{j1} are \code{Fourvector()} classes. Specifically, they are contravariant vectors $(E, \vec p)$. Their elements are accessed by \eg \code{E = j0[0]} and \code{p\_x = j0[1]}. Such objects can be added in a simple manner. 
\begin{rawcode}
from LHCO_reader import LHCO_reader
events = LHCO_reader.Events(f_name="example.lhco")

# Look at four-momentum of two jets in zeroth event
j0 = events[0]["jet"][0].vector()
j1 = events[0]["jet"][1].vector()

p = j0 + j1
\end{rawcode}
\begin{cliprawcode}
&\command& p = j0 + j1
\end{cliprawcode}
We can consider the invariant mass of such momenta:
\begin{rawcode}
from LHCO_reader import LHCO_reader
events = LHCO_reader.Events(f_name="example.lhco")

# Look at four-momentum of two jets in zeroth event
j0 = events[0]["jet"][0].vector()
j1 = events[0]["jet"][1].vector()

p = j0 + j1
m = abs(p)
\end{rawcode}
\begin{cliprawcode}
&\command& m = abs(p)
\end{cliprawcode}
This is equivalent to \code{(p**2)**0.5} and \code{(p * p)**0.5}, \ie one can perform Minkowski-squares and products with a mostly-minus metric, \eg
\begin{rawcode}
from LHCO_reader import LHCO_reader
events = LHCO_reader.Events(f_name="example.lhco")

# Look at four-momentum of two jets in zeroth event
j0 = events[0]["jet"][0].vector()
j1 = events[0]["jet"][1].vector()

product = j0 * j1
\end{rawcode}
\begin{cliprawcode}
&\command& product = j0 * j1
\end{cliprawcode}
One can even boost four-momenta. For example, we can find the boost to the \code{j0} rest-frame and boost \code{j1} and \code{j0} into that frame:
\begin{rawcode}
from LHCO_reader import LHCO_reader
events = LHCO_reader.Events(f_name="example.lhco")

# Look at four-momentum of two jets in zeroth event
j0 = events[0]["jet"][0].vector()
j1 = events[0]["jet"][1].vector()

beta = j0.beta_rest()
j0_prime = j0.boost(beta)
j1_prime = j1.boost(beta)
\end{rawcode}
\begin{cliprawcode}
&\command& beta = j0.beta_rest()
&\command& j0_prime = j0.boost(beta)
&\command& j1_prime = j1.boost(beta)
\end{cliprawcode}
The first line finds a three-vector describing a boost into the rest-frame of the first jet. The remaining lines boost the jets into that frame. The argument \code{beta} could have been any three-vector $\vec\beta$ describing a boost. We can make complicated cuts with four-momentum:
\begin{rawcode}
from LHCO_reader import LHCO_reader
events = LHCO_reader.Events(f_name="example.lhco")

import itertools
def W_boson(event):
    # Look at all pairs of jets
    for j1, j2 in itertools.combinations(event["jet"], 2):
        j1 = j1.vector()
        j2 = j2.vector()
        j = j1 + j2
        # Find invariant mass of pair
        M_W = abs(j)
        # Check if it agrees with W-boson
        if abs(M_W - 80.) < 10.:
            return True
    # If reached here, no W-bosons found
    return False

events.cut(W_boson)
\end{rawcode}

\begin{cliprawcode}
&\command& import itertools
&\command& def W_boson(event):
&\command&     # Look at all pairs of jets
&\command&     for j1, j2 in itertools.combinations(event["jet"], 2):
&\command&         j1 = j1.vector()
&\command&         j2 = j2.vector()
&\command&         j = j1 + j2
&\command&         # Find invariant mass of pair
&\command&         M_W = abs(j) 
&\command&         # Check if it agrees with W-boson
&\command&         if abs(M_W - 80.) < 10.:
&\command&             return True       
&\command&     # If reached here, no W-bosons found    
&\command&     return False
&\command&  
&\command& events.cut(W_boson) 
\end{cliprawcode}
This code cuts events without jets consistent with a $W$-boson's hadronic decay, based upon the invariant mass of pairs of jets.

One can also build an object and inspect the four-momenta, without an \LHCO file:
\begin{rawcode}
from LHCO_reader import LHCO_reader

electron = LHCO_reader.Object()
electron["PT"] = 10
electron["eta"] = 1
electron["phi"] = 1
electron["jmass"] = 0.
print electron.vector() 
\end{rawcode}
\begin{cliprawcode}
&\command& electron = LHCO_reader.Object()
&\command& electron["PT"] = 10
&\command& electron["eta"] = 1
&\command& electron["phi"] = 1
&\command& electron["jmass"] = 0.
&\command& print electron.vector() 
+---------------+---------------+---------------+---------------+
|       E       |      P_x      |      P_y      |      P_z      |
+---------------+---------------+---------------+---------------+
| 15.4308063482 | 5.40302305868 | 8.41470984808 | 11.7520119364 |
+---------------+---------------+---------------+---------------+
\end{cliprawcode}

\subsection{Cutting objects from an event}
As well as cutting events from the list of events, it is often useful to cut objects from an event. You might, for example, wish to discard any soft jets. This is achieved by \code{cut\_objects()} functions. The \code{Events()} and \code{Objects()} classes contain \code{cut\_objects()} functions for removing objects from all events, and for removing objects in a particular event, respectively. For example, to remove all soft jets from all events:
\begin{rawcode}
from LHCO_reader import LHCO_reader
events = LHCO_reader.Events(f_name="example.lhco")

PT = lambda _object: _object["PT"] < 30.
events.cut_objects("jet", PT)
\end{rawcode}
\begin{cliprawcode}
&\command& PT = lambda _object: _object["PT"] < 30.
&\command& events.cut_objects("jet", PT)
\end{cliprawcode}
The first line defines a function that returns \code{True} only if the object should be removed. The second line cuts all jets for which the function returns \code{True}.

\subsection{Quick plots}
After cutting away events, you might want to make a quick plot. This is possible with \code{matplotlib}. You can, of course, write your own \code{matplotlib} codes, but for a quick plot
\begin{rawcode}
from LHCO_reader import LHCO_reader
events = LHCO_reader.Events(f_name="example.lhco")

events.plot("electron", "PT")
\end{rawcode}
\begin{cliprawcode}
&\command& events.plot("electron", "PT")
&\includegraphics[width=0.7\textwidth]{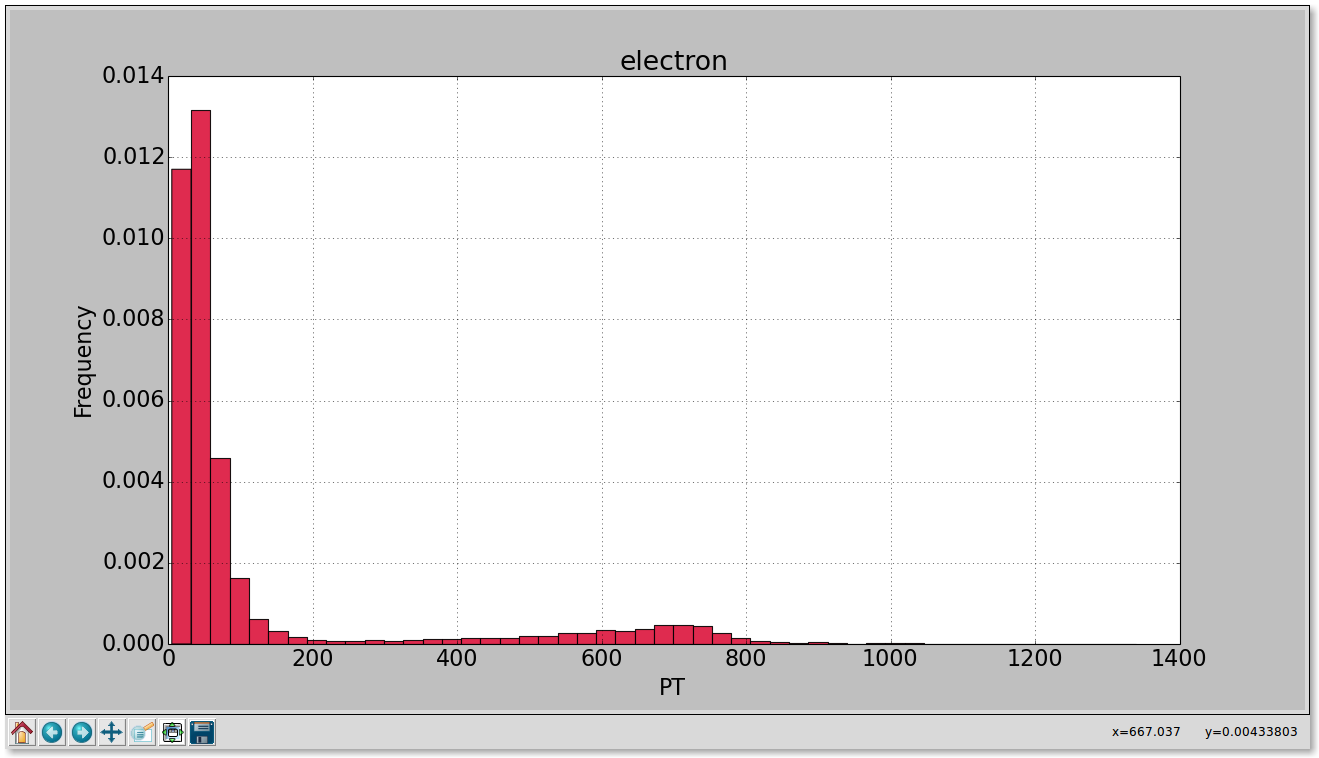}&
\end{cliprawcode}
You should see a histogram of the transverse momentum of all electrons, as displayed above. You could have picked any object in place of \code{electron} and any property in place of \code{PT}.

\subsection{Simple kinematics and angular separation}
The longitudinally-boost invariant angular separation, $\Delta R$, is common in analyses:
\begin{equation}
\Delta R = \sqrt{\Delta \phi^2 + \Delta \eta^2}.
\end{equation}
You can calculate it with the \code{delta\_R()} function:
\begin{rawcode}
from LHCO_reader import LHCO_reader
events = LHCO_reader.Events(f_name="example.lhco")

j0 = events[0]["jet"][0]
j1 = events[0]["jet"][1]
delta_R = LHCO_reader.delta_R(j0, j1)
\end{rawcode}
\begin{cliprawcode}
&\command& j0 = events[0]["jet"][0]
&\command& j1 = events[0]["jet"][1]
&\command& delta_R = LHCO_reader.delta_R(j0, j1)
\end{cliprawcode}
This code calculates $\Delta R$ for the two jets in the zeroth event.

Every event also includes a function for total transverse energy (\code{ET()}), hadronic transverse energy (\code{HT()}), missing
transverse energy (\code{MET()}) and missing hadronic energy (\code{MHT()}), with standard definitions, \eg
\begin{rawcode}
from LHCO_reader import LHCO_reader
events = LHCO_reader.Events(f_name="example.lhco")

print events[0].MHT()
print events[0].MHT()
print events[0].ET()
print events[0].MET()
\end{rawcode}
\begin{cliprawcode}
&\command& print events[0].HT()
&\command& print events[0].MHT()
&\command& print events[0].ET()
&\command& print events[0].MET()
\end{cliprawcode}
As noted in \refcite{Conte:2014xya}, detector-simulators compute missing transverse energy (MET) with an internal definition, that might differ from that inside \LHCOreader. The \LHCOreader definitions are that 
\begin{align}
E_T \equiv& \sum_{\text{Visible}} \left| \vec p_T\right|\\
\slashed{E}_T \equiv& \left|\sum_{\text{Visible}} \vec p_T\right|\\
H_T \equiv& \sum_{\text{Jets}} \left| \vec p_T\right|\\
\slashed{H}_T \equiv& \left|\sum_{\text{Jets}} \vec p_T\right|
\end{align}
By default, \LHCOreader calculates MET itself, though by supplying a keyword in the function call, \ie \code{MET(LHCO=True)}, it reads MET from the \LHCO file.

\subsection{Complicated kinematic variables}

\LHCOreader also includes complicated kinematic variables, including $\alpha_T$, razor, $M_{T2}$ and $M_{T2}^W$. The latter are calculated by Python interfaces to external libraries. 

\subsubsection{$\alpha_T$ and razor variables}
The $\alpha_T$\cite{Randall:2008rw,Khachatryan:2011tk} and razor\cite{Rogan:2010kb,Chatrchyan:2011ek} variables require that jets are arranged into mega-jets or pseudo-jets. This is achieved with a module of functions, \code{partition\_problem}, implementing algorithms to solve a partition problem. For $\alpha_T$, jets are arranged into pseudo-jets such that 
\begin{equation}
\Delta H_T = \left|\sum_{j \in j_1} H_T - \sum_{j \in j_2} H_T\right|
\end{equation}
is minimized. This is an optimization version of the partition problem, in which one must arrange a set into two equal subsets. We implement two incomplete algorithms: a greedy algorithm (\code{greedy}) and Karmarkar-Karp\cite{KK} (\code{KK}); and two complete algorithms: brute force (\code{brute}) and complete Karmarkar-Karp\cite{Korf98acomplete} (\code{CKK}). If one requires a quick approximate solution, Karmarkar-Karp is more precise than a greedy algorithm of identical complexity, $\mathcal{O}(n)$, whereas if one requires an optimal solution, complete Karmarkar-Karp is faster than brute force. Although the complete Karmarkar-Karp and brute force algorithms have equal complexity, $\mathcal{O}(2^n)$, complete Karmarkar-Karp prunes branches that could not yield an improvement, resulting in a speed-up. To our knowledge, at present only brute-force algorithms are implemented in public codes for arranging jets into mega-jets or pseudo-jets.

The $\alpha_T$ variable is calculated in the usual manner,
\begin{equation}
\alpha_T = \frac12 \frac{H_T - \Delta H_T}{\sqrt{H_T^2 - \Delta H_T^2}}
\end{equation}
by \eg \code{event[0].alpha\_T()}. The algorithm is set by \eg \code{LHCO\_reader.ALPHA\_T\_ALGORITHM = "KK"}. The default algorithm is \code{CKK}. The $\alpha_T$ variable was cross-checked with \code{oxbridge\_kinetics} in several examples in \code{kinetic\_test}.

The razor variable requires that jets are clustered into mega-jets such that the sum of the invariant masses 
\begin{equation}
m = \left|\sum_{j \in j_1} j^\mu\right| + \left|\sum_{j \in j_2} j^\mu\right|
\end{equation}
is minimized. This is an unusual variant of a partition problem. We know of no particularly good algorithms, but implement a brute force algorithm (\code{non\_standard\_brute}) and a greedy algorithm (\code{non\_standard\_greedy}). Whilst the greedy algorithm is fast, $\mathcal{O}(n)$, it might not be sufficiently accurate in all cases, though appears to be accurate to within a few percent in our tests. The razor variables are calculated in the usual manner:
\begin{align}
M^R &= \sqrt{(|\vec j_1| + |\vec j_2|)^2 - (j_1^z j_2^z)^2}\\
M^R_T &= \sqrt{1/2} \sqrt{\slashed{p}_T j_T - \slashed{p}_x j_x - \slashed{p}_y j_y}\\
R &= M_T^R / M_R
\end{align}
where $j \equiv j_1 + j_2$, the sum of the four-momenta of the two mega-jets, and $\slashed{p}$ is missing momenta. The variables are calculated by attributes of the \code{Event} class, \ie \code{event[0].razor\_MR()}, \code{event[0].razor\_MRT()} and \code{event[0].razor\_R()}. The algorithm for mega-jets is set by \eg \code{LHCO\_reader.RAZOR\_ALGORITHM = "non\_standard\_greedy"}. The default algorithm is \code{non\_standard\_brute}.

\subsubsection{$M_{T2}$ variables}
The $M_{T2}$\cite{Lester:2007fq,Barr:2003rg,Lester:1999tx} and $M_{T2}^W$\cite{Bai:2012gs} variables are implemented by \code{scipy.weave} interfaces to \code{C++} programs in \code{oxbridge\_kinetics.py} and \code{mt2w.py}, respectively. This requires the Oxbridge kinetics library developed by Lester et al\cite{oxbridge_kinetics}, which itself utilized an algorithm by Cheng and Han\cite{Cheng:2008hk}, and an $M_{T2}^W$ code by Gu et al\cite{mt2w}. The codes must be built following their own instructions, but for \code{mt2w}, you must in addition build a library: \code{gcc -shared -o libmt2w.so -fPIC mt2w\_bisect.cpp}. The paths to the libraries must be declared in \code{bash} variables:
\begin{lstlisting}
export MT2W=YOUR/PATH/TO/M2TW/CODE/
export OXBRIDGE=YOUR/PATH/TO/OXBRIDGE/CODE
export OXBRIDGE_LIB=YOUR/PATH/TO/OXBRIDGE/LIBRARY/
\end{lstlisting}
That is, \code{OXBRIDGE\_LIB} is a directory containing \code{liboxbridgekinetics-1.0.so}, whereas \code{OXBRIDGE} should contain \code{Mt2/ChengHanBisect\_Mt2\_332\_Calculator.h}, and \code{MT2W} should contain \code{mt2w\_bisect.h} and \code{libmt2w.so}.

The modules \code{oxbridge\_kinetics.py} and \code{mt2w.py} provide functions, \ie not attributes of any \LHCOreader classes. They are utilized by \eg
\begin{rawcode}
from LHCO_reader import LHCO_reader
from LHCO_reader import mt2w
from LHCO_reader import oxbridge_kinetics

events = LHCO_reader.Events(f_name="example.lhco")

lepton = events[0]["electron"][0]
jet_1 = events[0]["jet"][0]
jet_2 = events[0]["jet"][1]
MET = events[0]["MET"][0]

print mt2w.MT2W(lepton, jet_1, jet_2, MET)
print oxbridge_kinetics.MT2(lepton, jet_1, MET)
\end{rawcode}
\begin{cliprawcode}
&\command& from LHCO_reader import mt2w
&\command& from LHCO_reader import oxbridge_kinetics
&\command& lepton = events[0]["electron"][0]
&\command& jet_1 = events[0]["jet"][0]
&\command& jet_2 = events[0]["jet"][1]
&\command& MET = events[0]["MET"][0]
&\command& print mt2w.MT2W(lepton, jet_1, jet_2, MET)
434.702514648
&\command& print oxbridge_kinetics.MT2(lepton, jet_1, MET)
18.5098501371
\end{cliprawcode}
Their arguments are \code{Object} classes of the objects under consideration for
the $M_{T2}$ and $M_{T2}^W$ variables; see the full documentation\cite{api} for
further information. The interfaces were checked in
\code{kinetic\_test.py} and \code{mt2w.py} by reproducing results in the 
example programs.

\subsection{Ordering objects in an event}
Often, you might wish to consider the order of objects in an event, \eg you might wish to order the jets by transverse momentum and select the hardest jets. This is readily achieved:
\begin{rawcode}
from LHCO_reader import LHCO_reader
events = LHCO_reader.Events(f_name="example.lhco")

jets = events[0]["jet"].order("PT")
hardest = jets[:2]
\end{rawcode}
\begin{cliprawcode}
&\command& jets = events[0]["jet"].order("PT")
&\command& hardest = jets[:2]
\end{cliprawcode}

The first line orders the jets in descending transverse momentum (any dictionary key was possible, \eg \code{eta}). The second line picks the hardest two events, \ie the first two events in the list of jets ordered by transverse momentum. The \code{order()} function sorts in place and returns a sorted list. This function includes an optional keyword \code{reversed}, \code{True} by default, with which you could specify the order in which objects should be sorted.

\subsection{Saving events in \LHCO format}
One can build a selection of events by either cutting events from an \code{Events()} class, as already described, or by appending events from an \LHCO file to a new \code{Events()} class:
\begin{rawcode}
from LHCO_reader import LHCO_reader

my_events = LHCO_reader.Events(description="Hand-picked events")
events = LHCO_reader.Events("example.lhco")
my_events += events[:100]  # Add first 100 events from LHCO file
\end{rawcode}
\begin{cliprawcode}
&\command& my_events = LHCO_reader.Events(description="Hand-picked events")
&\command& events = LHCO_reader.Events("example.lhco")
&\command& my_events += events[:100]  # Add first 100 events from LHCO file
\end{cliprawcode}
Having made any selections, you can save your events in \LHCO format:
\begin{rawcode}
from LHCO_reader import LHCO_reader

my_events = LHCO_reader.Events(description="Hand-picked events")
events = LHCO_reader.Events("example.lhco")
my_events += events[:100]  # Add first 100 events from LHCO file
my_events.LHCO("my_events.lhco")
\end{rawcode}
\begin{cliprawcode}
&\command& my_events.LHCO("my_events.lhco")
\end{cliprawcode}
This writes an \LHCO file named \code{my\_events.lhco}.

\subsection{Compatibility with \ROOT}
\LHCOreader reads and writes \ROOT files by invoking \code{root2lhco} or \code{lhco2root} in \code{Delphes}; it cannot read and write \ROOT files by itself. To use \ROOT files, you must export the relevant \code{bash} variable to your \code{Delphes} build,
\begin{lstlisting}
export DELPHES=YOUR/PATH/TO/DELPHES
\end{lstlisting}
Having done so, you can read and write \ROOT files:
\begin{rawcode}
from LHCO_reader import LHCO_reader

events = LHCO_reader.Events(f_name="example.lhco")
events.ROOT("my_events.root")  # Write my_events.root
events = LHCO_reader.Events("my_events.root")  # Read my_events.root
events.ROOT("my_events.root")  # Write my_events.root
\end{rawcode}
\begin{cliprawcode}
&\command& events = LHCO_reader.Events("my_events.root")  # Read my_events.root
&\command& events.ROOT("my_events.root")  # Write my_events.root
\end{cliprawcode}
This reads \ROOT file named \code{my\_events.root} via an intermediate \LHCO file and writes an identical \ROOT file via an intermediate \LHCO file. The functions invoking \code{Delphes} are in the module \code{LHCO\_converter}. \LHCOreader is functional without \code{Delphes} --- only functions in \code{LHCO\_converter} fail without \code{Delphes}.

\subsection{Making an analysis}
There are many possible strategies for implementing an analysis. One could use the \code{cut()} function with an iterator of user-defined cuts:
\begin{rawcode}
from LHCO_reader import LHCO_reader
events = LHCO_reader.Events(f_name="example.lhco")

tau_cut = lambda event: event.number()["tau"] == 1
jet_cut = lambda event: event.number()["jet"] == 2
my_cuts = (tau_cut, jet_cut)

for my_cut in my_cuts:
    events.cut(my_cut)

print len(events)
\end{rawcode}
\begin{cliprawcode}
&\command& tau_cut = lambda event: event.number()["tau"] == 1
&\command& jet_cut = lambda event: event.number()["jet"] == 2
&\command& my_cuts = (tau_cut, jet_cut)
&\command& for my_cut in my_cuts:
&\command&     events.cut(my_cut)
&\command& 
&\command& print len(events)
\end{cliprawcode}
The initial commands define and apply two cuts. The final command prints the number of remaining events. Alternatively, one could loop over events, ignoring events that fail a cut, and count the number of acceptable events:
\begin{rawcode}
from LHCO_reader import LHCO_reader
events = LHCO_reader.Events(f_name="example.lhco")

tau_cut = lambda event: event.number()["tau"] == 1
jet_cut = lambda event: event.number()["jet"] == 2

n = 0 
for event in events:
    if tau_cut(event): continue
    if jet_cut(event): continue
    # Passed all cuts; increment counter
    n += 1
\end{rawcode}
\begin{cliprawcode}
&\command& n = 0 
&\command& for event in events:
&\command&     if tau_cut(event): continue
&\command&     if jet_cut(event): continue
&\command&     # Passed all cuts; increment counter
&\command&     n += 1
\end{cliprawcode}
The only point at which care must be taken is removing events from the list-like \code{Events()} class. Removing an element from a list shifts values of remaining elements in the list, which can cause surprises, \eg:
\begin{lstlisting}
&\command& example = [1.0, 2.0, 3.0, 4.0, 5.0, 6.0]
&\command& for e in example:
&\command&     if e < 3.:
&\command&         example.remove(e)
&\command& print example 
[2.0, 3.0, 4.0, 5.0, 6.0]
\end{lstlisting}
The \code{for} loop begins at the index \code{0} element in the \code{example} list. That element is removed, as it corresponds to a value of \code{1.} which is less than three. After which, the values shift such that index \code{0} corresponds to the value \code{2.}, which previously appeared at index \code{1}. The \code{for} loop proceeds to index \code{1} missing and never removing \code{2.} from the \code{example} list. This is overcome by making copies of the list
\begin{lstlisting}
&\command& import copy
&\command& example = [1.0, 2.0, 3.0, 4.0, 5.0, 6.0]
&\command& copy_example = copy.deepcopy(example)
&\command& for e in copy_example:
&\command&     if e < 3.:
&\command&         example.remove(e)
&\command& print example 
[3.0, 4.0, 5.0, 6.0]
\end{lstlisting}
or by reversing the order of the \code{for} loop:
\begin{lstlisting}
&\command& example = [1.0, 2.0, 3.0, 4.0, 5.0, 6.0]
&\command& for e in reversed(example):
&\command&     if e < 3.:
&\command&         example.remove(e)
&\command& print example 
[3.0, 4.0, 5.0, 6.0]
\end{lstlisting}

\section{Complete example}\label{Sec:Example}
We present an example analysis --- \attachfile[appearance=false,icon=Paperclip,mimetype=text/tex]{Example.py} --- similar to that in \refcite{Fowlie:2014iua}, in which we search for a $W^\prime$ gauge boson produced at the LHC and decaying via the chain
\beq
\label{Eq:DecayChain}
pp \to W^\prime \to e \nu_e^R  \to e e W^{\prime*} \to e e j j,
\eeq
and with a mass of about $2\tev$. This analysis mimics a CMS search\cite{Khachatryan:2014dka}. In our example analysis, we impose the following selections on detector-level jets and electrons:
\begin{enumerate}
\item\label{Item:Veto_ETA} Veto jets and electrons with pseudo-rapidity $\eta>2.5$.
\item\label{Item:Veto_PT} Veto jets with $P_T<40\gev$.
\item\label{Item:Electron_Number} Require exactly two electrons (of any charge).
\item Require at least two jets with $P_T>40\gev$.
\item Require that the hardest electron had transverse momentum $P_T>60\gev$ and the second hardest electron had $P_T>40\gev$.
\item Require that the invariant mass of the two hardest electrons was $M_{ee} > 200\gev$.
\item\label{Item:Invariant_Mass} Require that the invariant mass of the two hardest electrons and the two hardest jets was $1800\gev< M_{eejj} < 2200\gev$.
\end{enumerate}

We first read the \code{example.lhco} events into an \code{Events()} class:
\begin{lstlisting}
&\command& # Read LHCO events from disk
&\command& from LHCO_reader import LHCO_reader
&\command& events = LHCO_reader.Events("example.lhco")
\end{lstlisting}
We define functions for the \code{cut\_objects()} function, with which we will veto jets and electrons with pseudo-rapidity $\eta>2.5$ (\refcut{Item:Veto_ETA}) and jets with $P_T<40\gev$ (\refcut{Item:Veto_PT}):
\begin{lstlisting}
&\command& # Define object cuts
&\command& eta = lambda _object: abs(_object["eta"]) > 2.5   
&\command& jet_PT = lambda _object: _object["PT"] < 40
\end{lstlisting}
The functions \code{eta()} and \code{jet\_PT()} return true only if the object should be discarded. Furthermore, we define functions for the \code{cut()} function, with which to cut events:
\begin{lstlisting}
&\command& # Cut events based on number of electrons
&\command& electron_number = lambda event: event.number()["electron"] < 2
&\command& 
&\command& # Cut events based on number of jets
&\command& jet_number = lambda event: event.number()["jet"] < 2
&\command& 
&\command& def electron_PT(event):
&\command&     """ Cut events based on electron P_T. """
&\command&     
&\command&     # Impose electron PT requirement
&\command&     event["electron"].order("PT")
&\command&     return event["electron"][0]["PT"] < 60. or event["electron"][1]["PT"] < 40.
&\command& 
&\command& def electron_invariant_mass(event):
&\command&     """ Cut events based on electron invariant mass. """
&\command& 
&\command&     # Find four-momentum of two hardest electrons
&\command&     event["electron"].order("PT")
&\command&     p_electron = event["electron"][0].vector() + event["electron"][1].vector()
&\command& 
&\command&     # Find invariant mass
&\command&     M_ll = abs(p_electron)
&\command& 
&\command&     return M_ll < 200
&\command& 
&\command& def invariant_mass(event):
&\command&     """ Cut events based on jet and electron invariant mass. """
&\command& 
&\command&     # Find four-momentum of two hardest jets
&\command&     event["jet"].order("PT")
&\command&     p_jet = event["jet"][0].vector() + event["jet"][1].vector()
&\command& 
&\command&     # Find four-momentum of two hardest electrons
&\command&     event["electron"].order("PT")
&\command&     p_electron = event["electron"][0].vector() + event["electron"][1].vector()
&\command& 
&\command&     # Sum four-vectors and find invariant mass
&\command&     p = p_jet + p_electron
&\command&     M = abs(p)
&\command&     
&\command&     return not 1800. < M < 2200.
\end{lstlisting}
The functions \code{electron\_number()}, \code{electron\_PT()}, \code{jet\_number()}, \code{electron\_invariant\_mass()}, \code{invariant\_mass()} correspond to \refcut{Item:Electron_Number} to \refcut{Item:Invariant_Mass} respectively.

Finally, we apply the cuts and vetoes to the events:
\begin{lstlisting}
&\command& # Apply event cuts
&\command& events.cut_objects("electron", eta)
&\command& events.cut_objects("jet", eta)
&\command& events.cut_objects("jet", jet_PT)
&\command& 
&\command& cuts = (electron_number, 
&\command&         electron_PT,
&\command&         jet_number, 
&\command&         electron_invariant_mass,
&\command&         invariant_mass
&\command&         )
&\command& 
&\command& for cut in cuts:
&\command&     events.cut(cut)
&\command& 
&\command& events.LHCO("cut_events.lhco")  # Save events
&\command& print events
\end{lstlisting}
In the final lines, we save the events in \LHCO format and print a summary to the screen:
\begin{lstlisting}[basicstyle=\scriptsize\tt]
+------------------+--------------+
| Number of events | 179          |
| Description      | example.lhco |
+------------------+--------------+

+---------------------------------------------------------------------------------+----------------------+
| eta = lambda _object: abs(_object["eta"]) > 2.5                                 | 1.0                  |
| eta = lambda _object: abs(_object["eta"]) > 2.5                                 | 1.0                  |
| jet_PT = lambda _object: _object["PT"] < 40                                     | 1.0                  |
| electron_number = lambda event: event.number()["electron"] < 2                  | 0.803                |
| def electron_PT(event):                                                         | 0.226400996264       |
|     """ Cut events based on electron P_T. """                                   |                      |
|                                                                                 |                      |
|     # Impose electron PT requirement                                            |                      |
|     event["electron"].order("PT")                                               |                      |
|     return event["electron"][0]["PT"] < 60. or event["electron"][1]["PT"] < 40. |                      |
| jet_number = lambda event: event.number()["jet"] < 2                            | 0.397139713971       |
| def electron_invariant_mass(event):                                             | 0.624653739612       |
|     """ Cut events based on electron invariant mass. """                        |                      |
|                                                                                 |                      |
|     # Find four-momentum of two hardest electrons                               |                      |
|     event["electron"].order("PT")                                               |                      |
|     p_electron = event["electron"][0].vector() + event["electron"][1].vector()  |                      |
|                                                                                 |                      |
|     # Find invariant mass                                                       |                      |
|     M_ll = abs(p_electron)                                                      |                      |
|                                                                                 |                      |
|     return M_ll < 200                                                           |                      |
| def invariant_mass(event):                                                      | 0.39689578714        |
|     """ Cut events based on jet and electron invariant mass. """                |                      |
|                                                                                 |                      |
|     # Find four-momentum of two hardest jets                                    |                      |
|     event["jet"].order("PT")                                                    |                      |
|     p_jet = event["jet"][0].vector() + event["jet"][1].vector()                 |                      |
|                                                                                 |                      |
|     # Find four-momentum of two hardest electrons                               |                      |
|     event["electron"].order("PT")                                               |                      |
|     p_electron = event["electron"][0].vector() + event["electron"][1].vector()  |                      |
|                                                                                 |                      |
|     # Sum four-vectors and find invariant mass                                  |                      |
|     p = p_jet + p_electron                                                      |                      |
|     M = abs(p)                                                                  |                      |
|                                                                                 |                      |
|     return not 1800. < M < 2200.                                                |                      |
| Combined acceptance                                                             | 0.0179               |
| 68% interval                                                                    | [0.01658, 0.01932]   |
+---------------------------------------------------------------------------------+----------------------+
\end{lstlisting}
We see the individual acceptances for our cuts and that the combined acceptance was about $2\%$.

\section{Performance}\label{Sec:Performance}
The slowest parts of \LHCOreader were identified with line-by-line profiling, and optimized where possible. As a result, loading an \LHCO file takes approximately $0.5\,\text{s}$ per $10\,\text{k}$ events on an ordinary computer:
\begin{clipcode}
python -m timeit -n 20 -s 'from LHCO_reader import LHCO_reader' \
'LHCO_reader.Events("example.lhco")'
20 loops, best of 3: 351 msec per loop
\end{clipcode}
though this depends on the number of objects per event. Cutting events is much faster: a simple cut, presented in the examples above, on the number of tau-leptons takes approximately $0.3\,\text{s}$ per $10\,\text{k}$ events on an ordinary laptop. The example analysis in \refsec{Sec:Example} takes approximately $0.5\,\text{s}$ per $10\,\text{k}$ events. Whilst this isn't particularly impressive, and likely much slower than \code{C++} backends in \code{Seer} and \code{MadAnalysis}, it is reasonable, running interactively, for $\lesssim 100\,\text{k}$ events, and many more if you wait a few minutes for the code to execute.

\section{Summary}\label{Sec:Summary}
We presented a Python module, \LHCOreader, for reading and analyzing detector-level events stored in \LHCO format. The emphasis is upon ease of ease, rather than performance, though the module is optimized where possible. A user is free to implement any analysis in Python, by accessing \LHCO data in a convenient manner and utilizing functions for cutting, ordering and inspecting events, including a four-momentum class. \LHCOreader is compatible with detector-level events from \code{PGS} in \LHCO format, and by immediate conversion to \LHCO, detector-level events from \code{Delphes} in \ROOT format.

\acknowledgements
Many thanks to Luca Marzola for suggestions and testing. This work in part was supported by the ARC Centre of Excellence for Particle Physics at the Tera-scale.

\bibliography{refs}

\begin{thebibliography}{51}%
\makeatletter
\providecommand \@ifxundefined [1]{%
 \@ifx{#1\undefined}
}%
\providecommand \@ifnum [1]{%
 \ifnum #1\expandafter \@firstoftwo
 \else \expandafter \@secondoftwo
 \fi
}%
\providecommand \@ifx [1]{%
 \ifx #1\expandafter \@firstoftwo
 \else \expandafter \@secondoftwo
 \fi
}%
\providecommand \natexlab [1]{#1}%
\providecommand \enquote  [1]{``#1''}%
\providecommand \bibnamefont  [1]{#1}%
\providecommand \bibfnamefont [1]{#1}%
\providecommand \citenamefont [1]{#1}%
\providecommand \href@noop [0]{\@secondoftwo}%
\providecommand \href [0]{\begingroup \@sanitize@url \@href}%
\providecommand \@href[1]{\@@startlink{#1}\@@href}%
\providecommand \@@href[1]{\endgroup#1\@@endlink}%
\providecommand \@sanitize@url [0]{\catcode `\\12\catcode `\$12\catcode
  `\&12\catcode `\#12\catcode `\^12\catcode `\_12\catcode `\%12\relax}%
\providecommand \@@startlink[1]{}%
\providecommand \@@endlink[0]{}%
\providecommand \url  [0]{\begingroup\@sanitize@url \@url }%
\providecommand \@url [1]{\endgroup\@href {#1}{\urlprefix }}%
\providecommand \urlprefix  [0]{URL }%
\providecommand \Eprint [0]{\href }%
\providecommand \doibase [0]{http://dx.doi.org/}%
\providecommand \selectlanguage [0]{\@gobble}%
\providecommand \bibinfo  [0]{\@secondoftwo}%
\providecommand \bibfield  [0]{\@secondoftwo}%
\providecommand \translation [1]{[#1]}%
\providecommand \BibitemOpen [0]{}%
\providecommand \bibitemStop [0]{}%
\providecommand \bibitemNoStop [0]{.\EOS\space}%
\providecommand \EOS [0]{\spacefactor3000\relax}%
\providecommand \BibitemShut  [1]{\csname bibitem#1\endcsname}%
\let\auto@bib@innerbib\@empty
\bibitem [{\citenamefont {Glashow}(1961)}]{Glashow:1961tr}%
  \BibitemOpen
  \bibfield  {author} {\bibinfo {author} {\bibfnamefont {S.}~\bibnamefont
  {Glashow}},\ }\href {\doibase 10.1016/0029-5582(61)90469-2} {\bibfield
  {journal} {\bibinfo  {journal} {Nucl.Phys.}\ }\textbf {\bibinfo {volume}
  {22}},\ \bibinfo {pages} {579} (\bibinfo {year} {1961})}\BibitemShut
  {NoStop}%
\bibitem [{\citenamefont {Salam}(1968)}]{salam:sm}%
  \BibitemOpen
  \bibfield  {author} {\bibinfo {author} {\bibfnamefont {A.}~\bibnamefont
  {Salam}},\ }in\ \href@noop {} {\emph {\bibinfo {booktitle} {Elementary
  Particle Theory: Relativistic Groups and Analyticity}}},\ \bibinfo {editor}
  {edited by\ \bibinfo {editor} {\bibfnamefont {N.}~\bibnamefont
  {Svartholm}}},\ \bibinfo {organization} {The Nobel Symposium 8}\ (\bibinfo
  {publisher} {Almqvist and Wiksell, Stockholm},\ \bibinfo {address}
  {Aspen\"asg\aa{}rden, G\"oteborg, Sweden},\ \bibinfo {year} {1968})\ pp.\
  \bibinfo {pages} {367--377}\BibitemShut {NoStop}%
\bibitem [{\citenamefont {Weinberg}(1967)}]{Weinberg:1967tq}%
  \BibitemOpen
  \bibfield  {author} {\bibinfo {author} {\bibfnamefont {S.}~\bibnamefont
  {Weinberg}},\ }\href {\doibase 10.1103/PhysRevLett.19.1264} {\bibfield
  {journal} {\bibinfo  {journal} {Phys.Rev.Lett.}\ }\textbf {\bibinfo {volume}
  {19}},\ \bibinfo {pages} {1264} (\bibinfo {year} {1967})}\BibitemShut
  {NoStop}%
\bibitem [{\citenamefont {Plehn}(2012)}]{Plehn:2009nd}%
  \BibitemOpen
  \bibfield  {author} {\bibinfo {author} {\bibfnamefont {T.}~\bibnamefont
  {Plehn}},\ }\href {\doibase 10.1007/978-3-642-24040-9} {\bibfield  {journal}
  {\bibinfo  {journal} {Lect. Notes Phys.}\ }\textbf {\bibinfo {volume}
  {844}},\ \bibinfo {pages} {1} (\bibinfo {year} {2012})},\ \Eprint
  {http://arxiv.org/abs/0910.4182} {arXiv:0910.4182 [hep-ph]} \BibitemShut
  {NoStop}%
\bibitem [{\citenamefont {Mangano}\ \emph {et~al.}(2003)\citenamefont
  {Mangano}, \citenamefont {Moretti}, \citenamefont {Piccinini}, \citenamefont
  {Pittau},\ and\ \citenamefont {Polosa}}]{Mangano:2002ea}%
  \BibitemOpen
  \bibfield  {author} {\bibinfo {author} {\bibfnamefont {M.~L.}\ \bibnamefont
  {Mangano}}, \bibinfo {author} {\bibfnamefont {M.}~\bibnamefont {Moretti}},
  \bibinfo {author} {\bibfnamefont {F.}~\bibnamefont {Piccinini}}, \bibinfo
  {author} {\bibfnamefont {R.}~\bibnamefont {Pittau}}, \ and\ \bibinfo {author}
  {\bibfnamefont {A.~D.}\ \bibnamefont {Polosa}},\ }\href {\doibase
  10.1088/1126-6708/2003/07/001} {\bibfield  {journal} {\bibinfo  {journal}
  {JHEP}\ }\textbf {\bibinfo {volume} {07}},\ \bibinfo {pages} {001} (\bibinfo
  {year} {2003})},\ \Eprint {http://arxiv.org/abs/hep-ph/0206293}
  {arXiv:hep-ph/0206293 [hep-ph]} \BibitemShut {NoStop}%
\bibitem [{\citenamefont {Alwall}\ \emph {et~al.}(2014)\citenamefont {Alwall},
  \citenamefont {Frederix}, \citenamefont {Frixione}, \citenamefont {Hirschi},
  \citenamefont {Maltoni}, \citenamefont {Mattelaer}, \citenamefont {Shao},
  \citenamefont {Stelzer}, \citenamefont {Torrielli},\ and\ \citenamefont
  {Zaro}}]{Alwall:2014hca}%
  \BibitemOpen
  \bibfield  {author} {\bibinfo {author} {\bibfnamefont {J.}~\bibnamefont
  {Alwall}}, \bibinfo {author} {\bibfnamefont {R.}~\bibnamefont {Frederix}},
  \bibinfo {author} {\bibfnamefont {S.}~\bibnamefont {Frixione}}, \bibinfo
  {author} {\bibfnamefont {V.}~\bibnamefont {Hirschi}}, \bibinfo {author}
  {\bibfnamefont {F.}~\bibnamefont {Maltoni}}, \bibinfo {author} {\bibfnamefont
  {O.}~\bibnamefont {Mattelaer}}, \bibinfo {author} {\bibfnamefont {H.~S.}\
  \bibnamefont {Shao}}, \bibinfo {author} {\bibfnamefont {T.}~\bibnamefont
  {Stelzer}}, \bibinfo {author} {\bibfnamefont {P.}~\bibnamefont {Torrielli}},
  \ and\ \bibinfo {author} {\bibfnamefont {M.}~\bibnamefont {Zaro}},\ }\href
  {\doibase 10.1007/JHEP07(2014)079} {\bibfield  {journal} {\bibinfo  {journal}
  {JHEP}\ }\textbf {\bibinfo {volume} {07}},\ \bibinfo {pages} {079} (\bibinfo
  {year} {2014})},\ \Eprint {http://arxiv.org/abs/1405.0301} {arXiv:1405.0301
  [hep-ph]} \BibitemShut {NoStop}%
\bibitem [{\citenamefont {Cafarella}\ \emph {et~al.}(2009)\citenamefont
  {Cafarella}, \citenamefont {Papadopoulos},\ and\ \citenamefont
  {Worek}}]{Cafarella:2007pc}%
  \BibitemOpen
  \bibfield  {author} {\bibinfo {author} {\bibfnamefont {A.}~\bibnamefont
  {Cafarella}}, \bibinfo {author} {\bibfnamefont {C.~G.}\ \bibnamefont
  {Papadopoulos}}, \ and\ \bibinfo {author} {\bibfnamefont {M.}~\bibnamefont
  {Worek}},\ }\href {\doibase 10.1016/j.cpc.2009.04.023} {\bibfield  {journal}
  {\bibinfo  {journal} {Comput. Phys. Commun.}\ }\textbf {\bibinfo {volume}
  {180}},\ \bibinfo {pages} {1941} (\bibinfo {year} {2009})},\ \Eprint
  {http://arxiv.org/abs/0710.2427} {arXiv:0710.2427 [hep-ph]} \BibitemShut
  {NoStop}%
\bibitem [{\citenamefont {Kilian}\ \emph {et~al.}(2011)\citenamefont {Kilian},
  \citenamefont {Ohl},\ and\ \citenamefont {Reuter}}]{Kilian:2007gr}%
  \BibitemOpen
  \bibfield  {author} {\bibinfo {author} {\bibfnamefont {W.}~\bibnamefont
  {Kilian}}, \bibinfo {author} {\bibfnamefont {T.}~\bibnamefont {Ohl}}, \ and\
  \bibinfo {author} {\bibfnamefont {J.}~\bibnamefont {Reuter}},\ }\href
  {\doibase 10.1140/epjc/s10052-011-1742-y} {\bibfield  {journal} {\bibinfo
  {journal} {Eur. Phys. J.}\ }\textbf {\bibinfo {volume} {C71}},\ \bibinfo
  {pages} {1742} (\bibinfo {year} {2011})},\ \Eprint
  {http://arxiv.org/abs/0708.4233} {arXiv:0708.4233 [hep-ph]} \BibitemShut
  {NoStop}%
\bibitem [{\citenamefont {Bolognesi}\ \emph {et~al.}(2012)\citenamefont
  {Bolognesi}, \citenamefont {Gao}, \citenamefont {Gritsan}, \citenamefont
  {Melnikov}, \citenamefont {Schulze}, \citenamefont {Tran},\ and\
  \citenamefont {Whitbeck}}]{Bolognesi:2012mm}%
  \BibitemOpen
  \bibfield  {author} {\bibinfo {author} {\bibfnamefont {S.}~\bibnamefont
  {Bolognesi}}, \bibinfo {author} {\bibfnamefont {Y.}~\bibnamefont {Gao}},
  \bibinfo {author} {\bibfnamefont {A.~V.}\ \bibnamefont {Gritsan}}, \bibinfo
  {author} {\bibfnamefont {K.}~\bibnamefont {Melnikov}}, \bibinfo {author}
  {\bibfnamefont {M.}~\bibnamefont {Schulze}}, \bibinfo {author} {\bibfnamefont
  {N.~V.}\ \bibnamefont {Tran}}, \ and\ \bibinfo {author} {\bibfnamefont
  {A.}~\bibnamefont {Whitbeck}},\ }\href {\doibase 10.1103/PhysRevD.86.095031}
  {\bibfield  {journal} {\bibinfo  {journal} {Phys. Rev.}\ }\textbf {\bibinfo
  {volume} {D86}},\ \bibinfo {pages} {095031} (\bibinfo {year} {2012})},\
  \Eprint {http://arxiv.org/abs/1208.4018} {arXiv:1208.4018 [hep-ph]}
  \BibitemShut {NoStop}%
\bibitem [{\citenamefont {Gao}\ \emph {et~al.}(2010)\citenamefont {Gao},
  \citenamefont {Gritsan}, \citenamefont {Guo}, \citenamefont {Melnikov},
  \citenamefont {Schulze},\ and\ \citenamefont {Tran}}]{Gao:2010qx}%
  \BibitemOpen
  \bibfield  {author} {\bibinfo {author} {\bibfnamefont {Y.}~\bibnamefont
  {Gao}}, \bibinfo {author} {\bibfnamefont {A.~V.}\ \bibnamefont {Gritsan}},
  \bibinfo {author} {\bibfnamefont {Z.}~\bibnamefont {Guo}}, \bibinfo {author}
  {\bibfnamefont {K.}~\bibnamefont {Melnikov}}, \bibinfo {author}
  {\bibfnamefont {M.}~\bibnamefont {Schulze}}, \ and\ \bibinfo {author}
  {\bibfnamefont {N.~V.}\ \bibnamefont {Tran}},\ }\href {\doibase
  10.1103/PhysRevD.81.075022} {\bibfield  {journal} {\bibinfo  {journal} {Phys.
  Rev.}\ }\textbf {\bibinfo {volume} {D81}},\ \bibinfo {pages} {075022}
  (\bibinfo {year} {2010})},\ \Eprint {http://arxiv.org/abs/1001.3396}
  {arXiv:1001.3396 [hep-ph]} \BibitemShut {NoStop}%
\bibitem [{\citenamefont {Schulze}\ and\ \citenamefont {Tran}(2015)}]{JHU}%
  \BibitemOpen
  \bibfield  {author} {\bibinfo {author} {\bibfnamefont {M.}~\bibnamefont
  {Schulze}}\ and\ \bibinfo {author} {\bibfnamefont {N.}~\bibnamefont {Tran}},\
  }\href@noop {} {}\bibinfo {howpublished} {\url{http://www.pha.jhu.edu/spin/}}
  (\bibinfo {year} {2015}),\ \bibinfo {note} {{Accessed: January
  2016}}\BibitemShut {NoStop}%
\bibitem [{\citenamefont {Gleisberg}\ and\ \citenamefont
  {Hoeche}(2008)}]{Gleisberg:2008fv}%
  \BibitemOpen
  \bibfield  {author} {\bibinfo {author} {\bibfnamefont {T.}~\bibnamefont
  {Gleisberg}}\ and\ \bibinfo {author} {\bibfnamefont {S.}~\bibnamefont
  {Hoeche}},\ }\href {\doibase 10.1088/1126-6708/2008/12/039} {\bibfield
  {journal} {\bibinfo  {journal} {JHEP}\ }\textbf {\bibinfo {volume} {12}},\
  \bibinfo {pages} {039} (\bibinfo {year} {2008})},\ \Eprint
  {http://arxiv.org/abs/0808.3674} {arXiv:0808.3674 [hep-ph]} \BibitemShut
  {NoStop}%
\bibitem [{\citenamefont {Paige}\ \emph {et~al.}(2003)\citenamefont {Paige},
  \citenamefont {Protopopescu}, \citenamefont {Baer},\ and\ \citenamefont
  {Tata}}]{Paige:2003mg}%
  \BibitemOpen
  \bibfield  {author} {\bibinfo {author} {\bibfnamefont {F.~E.}\ \bibnamefont
  {Paige}}, \bibinfo {author} {\bibfnamefont {S.~D.}\ \bibnamefont
  {Protopopescu}}, \bibinfo {author} {\bibfnamefont {H.}~\bibnamefont {Baer}},
  \ and\ \bibinfo {author} {\bibfnamefont {X.}~\bibnamefont {Tata}},\
  }\href@noop {} {\  (\bibinfo {year} {2003})},\ \Eprint
  {http://arxiv.org/abs/hep-ph/0312045} {arXiv:hep-ph/0312045 [hep-ph]}
  \BibitemShut {NoStop}%
\bibitem [{\citenamefont {Pukhov}\ \emph {et~al.}(1999)\citenamefont {Pukhov},
  \citenamefont {Boos}, \citenamefont {Dubinin}, \citenamefont {Edneral},
  \citenamefont {Ilyin}, \citenamefont {Kovalenko}, \citenamefont {Kryukov},
  \citenamefont {Savrin}, \citenamefont {Shichanin},\ and\ \citenamefont
  {Semenov}}]{Pukhov:1999gg}%
  \BibitemOpen
  \bibfield  {author} {\bibinfo {author} {\bibfnamefont {A.}~\bibnamefont
  {Pukhov}}, \bibinfo {author} {\bibfnamefont {E.}~\bibnamefont {Boos}},
  \bibinfo {author} {\bibfnamefont {M.}~\bibnamefont {Dubinin}}, \bibinfo
  {author} {\bibfnamefont {V.}~\bibnamefont {Edneral}}, \bibinfo {author}
  {\bibfnamefont {V.}~\bibnamefont {Ilyin}}, \bibinfo {author} {\bibfnamefont
  {D.}~\bibnamefont {Kovalenko}}, \bibinfo {author} {\bibfnamefont
  {A.}~\bibnamefont {Kryukov}}, \bibinfo {author} {\bibfnamefont
  {V.}~\bibnamefont {Savrin}}, \bibinfo {author} {\bibfnamefont
  {S.}~\bibnamefont {Shichanin}}, \ and\ \bibinfo {author} {\bibfnamefont
  {A.}~\bibnamefont {Semenov}},\ }\href@noop {} {\  (\bibinfo {year} {1999})},\
  \Eprint {http://arxiv.org/abs/hep-ph/9908288} {arXiv:hep-ph/9908288 [hep-ph]}
  \BibitemShut {NoStop}%
\bibitem [{\citenamefont {Gleisberg}\ \emph {et~al.}(2009)\citenamefont
  {Gleisberg}, \citenamefont {Hoeche}, \citenamefont {Krauss}, \citenamefont
  {Schonherr}, \citenamefont {Schumann}, \citenamefont {Siegert},\ and\
  \citenamefont {Winter}}]{Gleisberg:2008ta}%
  \BibitemOpen
  \bibfield  {author} {\bibinfo {author} {\bibfnamefont {T.}~\bibnamefont
  {Gleisberg}}, \bibinfo {author} {\bibfnamefont {S.}~\bibnamefont {Hoeche}},
  \bibinfo {author} {\bibfnamefont {F.}~\bibnamefont {Krauss}}, \bibinfo
  {author} {\bibfnamefont {M.}~\bibnamefont {Schonherr}}, \bibinfo {author}
  {\bibfnamefont {S.}~\bibnamefont {Schumann}}, \bibinfo {author}
  {\bibfnamefont {F.}~\bibnamefont {Siegert}}, \ and\ \bibinfo {author}
  {\bibfnamefont {J.}~\bibnamefont {Winter}},\ }\href {\doibase
  10.1088/1126-6708/2009/02/007} {\bibfield  {journal} {\bibinfo  {journal}
  {JHEP}\ }\textbf {\bibinfo {volume} {02}},\ \bibinfo {pages} {007} (\bibinfo
  {year} {2009})},\ \Eprint {http://arxiv.org/abs/0811.4622} {arXiv:0811.4622
  [hep-ph]} \BibitemShut {NoStop}%
\bibitem [{\citenamefont {Sjostrand}\ \emph {et~al.}(2008)\citenamefont
  {Sjostrand}, \citenamefont {Mrenna},\ and\ \citenamefont
  {Skands}}]{Sjostrand:2007gs}%
  \BibitemOpen
  \bibfield  {author} {\bibinfo {author} {\bibfnamefont {T.}~\bibnamefont
  {Sjostrand}}, \bibinfo {author} {\bibfnamefont {S.}~\bibnamefont {Mrenna}}, \
  and\ \bibinfo {author} {\bibfnamefont {P.~Z.}\ \bibnamefont {Skands}},\
  }\href {\doibase 10.1016/j.cpc.2008.01.036} {\bibfield  {journal} {\bibinfo
  {journal} {Comput.Phys.Commun.}\ }\textbf {\bibinfo {volume} {178}},\
  \bibinfo {pages} {852} (\bibinfo {year} {2008})},\ \Eprint
  {http://arxiv.org/abs/0710.3820} {arXiv:0710.3820 [hep-ph]} \BibitemShut
  {NoStop}%
\bibitem [{\citenamefont {Bahr}\ \emph {et~al.}(2008)\citenamefont {Bahr} \emph
  {et~al.}}]{Bahr:2008pv}%
  \BibitemOpen
  \bibfield  {author} {\bibinfo {author} {\bibfnamefont {M.}~\bibnamefont
  {Bahr}} \emph {et~al.},\ }\href {\doibase 10.1140/epjc/s10052-008-0798-9}
  {\bibfield  {journal} {\bibinfo  {journal} {Eur. Phys. J.}\ }\textbf
  {\bibinfo {volume} {C58}},\ \bibinfo {pages} {639} (\bibinfo {year}
  {2008})},\ \Eprint {http://arxiv.org/abs/0803.0883} {arXiv:0803.0883
  [hep-ph]} \BibitemShut {NoStop}%
\bibitem [{\citenamefont {Lonnblad}(1992)}]{Lonnblad:1992tz}%
  \BibitemOpen
  \bibfield  {author} {\bibinfo {author} {\bibfnamefont {L.}~\bibnamefont
  {Lonnblad}},\ }\href {\doibase 10.1016/0010-4655(92)90068-A} {\bibfield
  {journal} {\bibinfo  {journal} {Comput. Phys. Commun.}\ }\textbf {\bibinfo
  {volume} {71}},\ \bibinfo {pages} {15} (\bibinfo {year} {1992})}\BibitemShut
  {NoStop}%
\bibitem [{\citenamefont {de~Favereau}\ \emph {et~al.}(2014)\citenamefont
  {de~Favereau}, \citenamefont {Delaere}, \citenamefont {Demin}, \citenamefont
  {Giammanco}, \citenamefont {Lemaître}, \citenamefont {Mertens},\ and\
  \citenamefont {Selvaggi}}]{deFavereau:2013fsa}%
  \BibitemOpen
  \bibfield  {author} {\bibinfo {author} {\bibfnamefont {J.}~\bibnamefont
  {de~Favereau}}, \bibinfo {author} {\bibfnamefont {C.}~\bibnamefont
  {Delaere}}, \bibinfo {author} {\bibfnamefont {P.}~\bibnamefont {Demin}},
  \bibinfo {author} {\bibfnamefont {A.}~\bibnamefont {Giammanco}}, \bibinfo
  {author} {\bibfnamefont {V.}~\bibnamefont {Lemaître}}, \bibinfo {author}
  {\bibfnamefont {A.}~\bibnamefont {Mertens}}, \ and\ \bibinfo {author}
  {\bibfnamefont {M.}~\bibnamefont {Selvaggi}} (\bibinfo {collaboration}
  {DELPHES 3}),\ }\href {\doibase 10.1007/JHEP02(2014)057} {\bibfield
  {journal} {\bibinfo  {journal} {JHEP}\ }\textbf {\bibinfo {volume} {02}},\
  \bibinfo {pages} {057} (\bibinfo {year} {2014})},\ \Eprint
  {http://arxiv.org/abs/1307.6346} {arXiv:1307.6346 [hep-ex]} \BibitemShut
  {NoStop}%
\bibitem [{\citenamefont {Conway}\ \emph {et~al.}(2012)\citenamefont {Conway}
  \emph {et~al.}}]{PGS}%
  \BibitemOpen
  \bibfield  {author} {\bibinfo {author} {\bibfnamefont {J.}~\bibnamefont
  {Conway}} \emph {et~al.},\ }\href@noop {} {\enquote {\bibinfo {title} {{PGS 4
  - Pretty Good Simulation of high-energy collisions}},}\ }\bibinfo
  {howpublished}
  {\url{http://www.physics.ucdavis.edu/~conway/research/software/pgs/pgs4-general.html}}
  (\bibinfo {year} {2012}),\ \bibinfo {note} {{Accessed: January
  2016}}\BibitemShut {NoStop}%
\bibitem [{\citenamefont {Thaler}(2006)}]{LHCO}%
  \BibitemOpen
  \bibfield  {author} {\bibinfo {author} {\bibfnamefont {J.}~\bibnamefont
  {Thaler}},\ }\href@noop {} {\enquote {\bibinfo {title} {{How to Read LHC
  Olympics Data Files}},}\ }\bibinfo {howpublished}
  {\url{http://madgraph.phys.ucl.ac.be/Manual/lhco.html}} (\bibinfo {year}
  {2006}),\ \bibinfo {note} {{Accessed: January 2016}}\BibitemShut {NoStop}%
\bibitem [{\citenamefont {Antcheva}\ \emph {et~al.}(2009)\citenamefont
  {Antcheva} \emph {et~al.}}]{Antcheva:2009zz}%
  \BibitemOpen
  \bibfield  {author} {\bibinfo {author} {\bibfnamefont {I.}~\bibnamefont
  {Antcheva}} \emph {et~al.},\ }\href {\doibase 10.1016/j.cpc.2009.08.005}
  {\bibfield  {journal} {\bibinfo  {journal} {Comput. Phys. Commun.}\ }\textbf
  {\bibinfo {volume} {180}},\ \bibinfo {pages} {2499} (\bibinfo {year}
  {2009})},\ \Eprint {http://arxiv.org/abs/1508.07749} {arXiv:1508.07749
  [physics.data-an]} \BibitemShut {NoStop}%
\bibitem [{\citenamefont {Conte}\ \emph {et~al.}(2013)\citenamefont {Conte},
  \citenamefont {Fuks},\ and\ \citenamefont {Serret}}]{Conte:2012fm}%
  \BibitemOpen
  \bibfield  {author} {\bibinfo {author} {\bibfnamefont {E.}~\bibnamefont
  {Conte}}, \bibinfo {author} {\bibfnamefont {B.}~\bibnamefont {Fuks}}, \ and\
  \bibinfo {author} {\bibfnamefont {G.}~\bibnamefont {Serret}},\ }\href
  {\doibase 10.1016/j.cpc.2012.09.009} {\bibfield  {journal} {\bibinfo
  {journal} {Comput. Phys. Commun.}\ }\textbf {\bibinfo {volume} {184}},\
  \bibinfo {pages} {222} (\bibinfo {year} {2013})},\ \Eprint
  {http://arxiv.org/abs/1206.1599} {arXiv:1206.1599 [hep-ph]} \BibitemShut
  {NoStop}%
\bibitem [{\citenamefont {Conte}\ and\ \citenamefont
  {Fuks}(2014)}]{Conte:2013mea}%
  \BibitemOpen
  \bibfield  {author} {\bibinfo {author} {\bibfnamefont {E.}~\bibnamefont
  {Conte}}\ and\ \bibinfo {author} {\bibfnamefont {B.}~\bibnamefont {Fuks}},\
  }\bibfield  {booktitle} {\emph {\bibinfo {booktitle} {{Proceedings, 15th
  International Workshop on Advanced Computing and Analysis Techniques in
  Physics Research (ACAT 2013)}}},\ }\href {\doibase
  10.1088/1742-6596/523/1/012032} {\bibfield  {journal} {\bibinfo  {journal}
  {J. Phys. Conf. Ser.}\ }\textbf {\bibinfo {volume} {523}},\ \bibinfo {pages}
  {012032} (\bibinfo {year} {2014})},\ \Eprint {http://arxiv.org/abs/1309.7831}
  {arXiv:1309.7831 [hep-ph]} \BibitemShut {NoStop}%
\bibitem [{\citenamefont {Conte}\ \emph {et~al.}(2015)\citenamefont {Conte},
  \citenamefont {Dumont}, \citenamefont {Fuks},\ and\ \citenamefont
  {Schmitt}}]{Conte:2014xya}%
  \BibitemOpen
  \bibfield  {author} {\bibinfo {author} {\bibfnamefont {E.}~\bibnamefont
  {Conte}}, \bibinfo {author} {\bibfnamefont {B.}~\bibnamefont {Dumont}},
  \bibinfo {author} {\bibfnamefont {B.}~\bibnamefont {Fuks}}, \ and\ \bibinfo
  {author} {\bibfnamefont {T.}~\bibnamefont {Schmitt}},\ }\bibfield
  {booktitle} {\emph {\bibinfo {booktitle} {{Proceedings, 16th International
  workshop on Advanced Computing and Analysis Techniques in physics (ACAT
  14)}}},\ }\href {\doibase 10.1088/1742-6596/608/1/012054} {\bibfield
  {journal} {\bibinfo  {journal} {J. Phys. Conf. Ser.}\ }\textbf {\bibinfo
  {volume} {608}},\ \bibinfo {pages} {012054} (\bibinfo {year} {2015})},\
  \Eprint {http://arxiv.org/abs/1410.2785} {arXiv:1410.2785 [hep-ph]}
  \BibitemShut {NoStop}%
\bibitem [{\citenamefont {Martin}(2015)}]{Martin:2015hra}%
  \BibitemOpen
  \bibfield  {author} {\bibinfo {author} {\bibfnamefont {T.~A.~W.}\
  \bibnamefont {Martin}},\ }\href@noop {} {\  (\bibinfo {year} {2015})},\
  \Eprint {http://arxiv.org/abs/1503.03073} {arXiv:1503.03073 [hep-ph]}
  \BibitemShut {NoStop}%
\bibitem [{\citenamefont {Walker}(2012)}]{Walker:2012vf}%
  \BibitemOpen
  \bibfield  {author} {\bibinfo {author} {\bibfnamefont {J.~W.}\ \bibnamefont
  {Walker}},\ }\href@noop {} {\  (\bibinfo {year} {2012})},\ \Eprint
  {http://arxiv.org/abs/1207.3383} {arXiv:1207.3383 [hep-ph]} \BibitemShut
  {NoStop}%
\bibitem [{\citenamefont {Conte}\ \emph {et~al.}(2014)\citenamefont {Conte},
  \citenamefont {Dumont}, \citenamefont {Fuks},\ and\ \citenamefont
  {Wymant}}]{Conte:2014zja}%
  \BibitemOpen
  \bibfield  {author} {\bibinfo {author} {\bibfnamefont {E.}~\bibnamefont
  {Conte}}, \bibinfo {author} {\bibfnamefont {B.}~\bibnamefont {Dumont}},
  \bibinfo {author} {\bibfnamefont {B.}~\bibnamefont {Fuks}}, \ and\ \bibinfo
  {author} {\bibfnamefont {C.}~\bibnamefont {Wymant}},\ }\href {\doibase
  10.1140/epjc/s10052-014-3103-0} {\bibfield  {journal} {\bibinfo  {journal}
  {Eur. Phys. J.}\ }\textbf {\bibinfo {volume} {C74}},\ \bibinfo {pages} {3103}
  (\bibinfo {year} {2014})},\ \Eprint {http://arxiv.org/abs/1405.3982}
  {arXiv:1405.3982 [hep-ph]} \BibitemShut {NoStop}%
\bibitem [{\citenamefont {Maurits}(2013)}]{pt}%
  \BibitemOpen
  \bibfield  {author} {\bibinfo {author} {\bibfnamefont {L.}~\bibnamefont
  {Maurits}},\ }\href@noop {} {\enquote {\bibinfo {title} {Prettytable
  0.7.2},}\ }\bibinfo {howpublished}
  {\url{https://pypi.python.org/pypi/PrettyTable}} (\bibinfo {year} {2013}),\
  \bibinfo {note} {{Accessed: January 2016}}\BibitemShut {NoStop}%
\bibitem [{\citenamefont {Walt}\ \emph {et~al.}(2011)\citenamefont {Walt},
  \citenamefont {Colbert},\ and\ \citenamefont {Varoquaux}}]{numpy}%
  \BibitemOpen
  \bibfield  {author} {\bibinfo {author} {\bibfnamefont {S.~v.~d.}\
  \bibnamefont {Walt}}, \bibinfo {author} {\bibfnamefont {S.~C.}\ \bibnamefont
  {Colbert}}, \ and\ \bibinfo {author} {\bibfnamefont {G.}~\bibnamefont
  {Varoquaux}},\ }\href {\doibase http://dx.doi.org/10.1109/MCSE.2011.37}
  {\bibfield  {journal} {\bibinfo  {journal} {Computing in Science \&
  Engineering}\ }\textbf {\bibinfo {volume} {13}},\ \bibinfo {pages} {22}
  (\bibinfo {year} {2011})}\BibitemShut {NoStop}%
\bibitem [{\citenamefont {Jones}\ \emph {et~al.}(01  )\citenamefont {Jones},
  \citenamefont {Oliphant}, \citenamefont {Peterson} \emph {et~al.}}]{scipy}%
  \BibitemOpen
  \bibfield  {author} {\bibinfo {author} {\bibfnamefont {E.}~\bibnamefont
  {Jones}}, \bibinfo {author} {\bibfnamefont {T.}~\bibnamefont {Oliphant}},
  \bibinfo {author} {\bibfnamefont {P.}~\bibnamefont {Peterson}},  \emph
  {et~al.},\ }\href@noop {} {\enquote {\bibinfo {title} {{{SciPy}: Open source
  scientific tools for {Python}}},}\ }\bibinfo {howpublished}
  {\url{http://www.scipy.org}} (\bibinfo {year} {2001--}),\ \bibinfo {note}
  {accessed: January 2016}\BibitemShut {NoStop}%
\bibitem [{\citenamefont {Millman}\ and\ \citenamefont
  {Aivazis}(2011)}]{scipy2}%
  \BibitemOpen
  \bibfield  {author} {\bibinfo {author} {\bibfnamefont {K.~J.}\ \bibnamefont
  {Millman}}\ and\ \bibinfo {author} {\bibfnamefont {M.}~\bibnamefont
  {Aivazis}},\ }\href {\doibase http://dx.doi.org/10.1109/MCSE.2011.36}
  {\bibfield  {journal} {\bibinfo  {journal} {Computing in Science \&
  Engineering}\ }\textbf {\bibinfo {volume} {13}},\ \bibinfo {pages} {9}
  (\bibinfo {year} {2011})}\BibitemShut {NoStop}%
\bibitem [{\citenamefont {Hunter}(2007)}]{matplotlib}%
  \BibitemOpen
  \bibfield  {author} {\bibinfo {author} {\bibfnamefont {J.}~\bibnamefont
  {Hunter}},\ }\href {\doibase 10.1109/MCSE.2007.55} {\bibfield  {journal}
  {\bibinfo  {journal} {Computing in Science Engineering}\ }\textbf {\bibinfo
  {volume} {9}},\ \bibinfo {pages} {90} (\bibinfo {year} {2007})}\BibitemShut
  {NoStop}%
\bibitem [{\citenamefont {Agresti}(2003)}]{agresti2003categorical}%
  \BibitemOpen
  \bibfield  {author} {\bibinfo {author} {\bibfnamefont {A.}~\bibnamefont
  {Agresti}},\ }\href {https://books.google.com.au/books?id=hpEzw4T0sPUC}
  {\emph {\bibinfo {title} {Categorical Data Analysis}}},\ Wiley Series in
  Probability and Statistics\ (\bibinfo  {publisher} {Wiley},\ \bibinfo {year}
  {2003})\BibitemShut {NoStop}%
\bibitem [{\citenamefont {CLOPPER}\ and\ \citenamefont
  {PEARSON}(1934)}]{CLOPPER01121934}%
  \BibitemOpen
  \bibfield  {author} {\bibinfo {author} {\bibfnamefont {C.~J.}\ \bibnamefont
  {CLOPPER}}\ and\ \bibinfo {author} {\bibfnamefont {E.~S.}\ \bibnamefont
  {PEARSON}},\ }\href {\doibase 10.1093/biomet/26.4.404} {\bibfield  {journal}
  {\bibinfo  {journal} {Biometrika}\ }\textbf {\bibinfo {volume} {26}},\
  \bibinfo {pages} {404} (\bibinfo {year} {1934})},\ \Eprint
  {http://arxiv.org/abs/http://biomet.oxfordjournals.org/content/26/4/404.full.pdf+html}
  {http://biomet.oxfordjournals.org/content/26/4/404.full.pdf+html}
  \BibitemShut {NoStop}%
\bibitem [{\citenamefont {Randall}\ and\ \citenamefont
  {Tucker-Smith}(2008)}]{Randall:2008rw}%
  \BibitemOpen
  \bibfield  {author} {\bibinfo {author} {\bibfnamefont {L.}~\bibnamefont
  {Randall}}\ and\ \bibinfo {author} {\bibfnamefont {D.}~\bibnamefont
  {Tucker-Smith}},\ }\href {\doibase 10.1103/PhysRevLett.101.221803} {\bibfield
   {journal} {\bibinfo  {journal} {Phys. Rev. Lett.}\ }\textbf {\bibinfo
  {volume} {101}},\ \bibinfo {pages} {221803} (\bibinfo {year} {2008})},\
  \Eprint {http://arxiv.org/abs/0806.1049} {arXiv:0806.1049 [hep-ph]}
  \BibitemShut {NoStop}%
\bibitem [{\citenamefont {Khachatryan}\ \emph {et~al.}(2011)\citenamefont
  {Khachatryan} \emph {et~al.}}]{Khachatryan:2011tk}%
  \BibitemOpen
  \bibfield  {author} {\bibinfo {author} {\bibfnamefont {V.}~\bibnamefont
  {Khachatryan}} \emph {et~al.} (\bibinfo {collaboration} {CMS}),\ }\href
  {\doibase 10.1016/j.physletb.2011.03.021} {\bibfield  {journal} {\bibinfo
  {journal} {Phys. Lett.}\ }\textbf {\bibinfo {volume} {B698}},\ \bibinfo
  {pages} {196} (\bibinfo {year} {2011})},\ \Eprint
  {http://arxiv.org/abs/1101.1628} {arXiv:1101.1628 [hep-ex]} \BibitemShut
  {NoStop}%
\bibitem [{\citenamefont {Rogan}(2010)}]{Rogan:2010kb}%
  \BibitemOpen
  \bibfield  {author} {\bibinfo {author} {\bibfnamefont {C.}~\bibnamefont
  {Rogan}},\ }\href@noop {} {\  (\bibinfo {year} {2010})},\ \Eprint
  {http://arxiv.org/abs/1006.2727} {arXiv:1006.2727 [hep-ph]} \BibitemShut
  {NoStop}%
\bibitem [{\citenamefont {Chatrchyan}\ \emph {et~al.}(2012)\citenamefont
  {Chatrchyan} \emph {et~al.}}]{Chatrchyan:2011ek}%
  \BibitemOpen
  \bibfield  {author} {\bibinfo {author} {\bibfnamefont {S.}~\bibnamefont
  {Chatrchyan}} \emph {et~al.} (\bibinfo {collaboration} {CMS}),\ }\href
  {\doibase 10.1103/PhysRevD.85.012004} {\bibfield  {journal} {\bibinfo
  {journal} {Phys. Rev.}\ }\textbf {\bibinfo {volume} {D85}},\ \bibinfo {pages}
  {012004} (\bibinfo {year} {2012})},\ \Eprint {http://arxiv.org/abs/1107.1279}
  {arXiv:1107.1279 [hep-ex]} \BibitemShut {NoStop}%
\bibitem [{\citenamefont {Karmarkar}\ and\ \citenamefont {Karp}(1982)}]{KK}%
  \BibitemOpen
  \bibfield  {author} {\bibinfo {author} {\bibfnamefont {N.}~\bibnamefont
  {Karmarkar}}\ and\ \bibinfo {author} {\bibfnamefont {R.~M.}\ \bibnamefont
  {Karp}},\ }\href@noop {} {\  (\bibinfo {year} {1982})}\BibitemShut {NoStop}%
\bibitem [{\citenamefont {Korf}(1998)}]{Korf98acomplete}%
  \BibitemOpen
  \bibfield  {author} {\bibinfo {author} {\bibfnamefont {R.~E.}\ \bibnamefont
  {Korf}},\ }\href@noop {} {\bibfield  {journal} {\bibinfo  {journal}
  {Artificial Intelligence}\ }\textbf {\bibinfo {volume} {106}},\ \bibinfo
  {pages} {181} (\bibinfo {year} {1998})}\BibitemShut {NoStop}%
\bibitem [{\citenamefont {Lester}\ and\ \citenamefont
  {Barr}(2007)}]{Lester:2007fq}%
  \BibitemOpen
  \bibfield  {author} {\bibinfo {author} {\bibfnamefont {C.}~\bibnamefont
  {Lester}}\ and\ \bibinfo {author} {\bibfnamefont {A.}~\bibnamefont {Barr}},\
  }\href {\doibase 10.1088/1126-6708/2007/12/102} {\bibfield  {journal}
  {\bibinfo  {journal} {JHEP}\ }\textbf {\bibinfo {volume} {12}},\ \bibinfo
  {pages} {102} (\bibinfo {year} {2007})},\ \Eprint
  {http://arxiv.org/abs/0708.1028} {arXiv:0708.1028 [hep-ph]} \BibitemShut
  {NoStop}%
\bibitem [{\citenamefont {Barr}\ \emph {et~al.}(2003)\citenamefont {Barr},
  \citenamefont {Lester},\ and\ \citenamefont {Stephens}}]{Barr:2003rg}%
  \BibitemOpen
  \bibfield  {author} {\bibinfo {author} {\bibfnamefont {A.}~\bibnamefont
  {Barr}}, \bibinfo {author} {\bibfnamefont {C.}~\bibnamefont {Lester}}, \ and\
  \bibinfo {author} {\bibfnamefont {P.}~\bibnamefont {Stephens}},\ }\href
  {\doibase 10.1088/0954-3899/29/10/304} {\bibfield  {journal} {\bibinfo
  {journal} {J. Phys.}\ }\textbf {\bibinfo {volume} {G29}},\ \bibinfo {pages}
  {2343} (\bibinfo {year} {2003})},\ \Eprint
  {http://arxiv.org/abs/hep-ph/0304226} {arXiv:hep-ph/0304226 [hep-ph]}
  \BibitemShut {NoStop}%
\bibitem [{\citenamefont {Lester}\ and\ \citenamefont
  {Summers}(1999)}]{Lester:1999tx}%
  \BibitemOpen
  \bibfield  {author} {\bibinfo {author} {\bibfnamefont {C.~G.}\ \bibnamefont
  {Lester}}\ and\ \bibinfo {author} {\bibfnamefont {D.~J.}\ \bibnamefont
  {Summers}},\ }\href {\doibase 10.1016/S0370-2693(99)00945-4} {\bibfield
  {journal} {\bibinfo  {journal} {Phys. Lett.}\ }\textbf {\bibinfo {volume}
  {B463}},\ \bibinfo {pages} {99} (\bibinfo {year} {1999})},\ \Eprint
  {http://arxiv.org/abs/hep-ph/9906349} {arXiv:hep-ph/9906349 [hep-ph]}
  \BibitemShut {NoStop}%
\bibitem [{\citenamefont {Bai}\ \emph {et~al.}(2012)\citenamefont {Bai},
  \citenamefont {Cheng}, \citenamefont {Gallicchio},\ and\ \citenamefont
  {Gu}}]{Bai:2012gs}%
  \BibitemOpen
  \bibfield  {author} {\bibinfo {author} {\bibfnamefont {Y.}~\bibnamefont
  {Bai}}, \bibinfo {author} {\bibfnamefont {H.-C.}\ \bibnamefont {Cheng}},
  \bibinfo {author} {\bibfnamefont {J.}~\bibnamefont {Gallicchio}}, \ and\
  \bibinfo {author} {\bibfnamefont {J.}~\bibnamefont {Gu}},\ }\href {\doibase
  10.1007/JHEP07(2012)110} {\bibfield  {journal} {\bibinfo  {journal} {JHEP}\
  }\textbf {\bibinfo {volume} {07}},\ \bibinfo {pages} {110} (\bibinfo {year}
  {2012})},\ \Eprint {http://arxiv.org/abs/1203.4813} {arXiv:1203.4813
  [hep-ph]} \BibitemShut {NoStop}%
\bibitem [{\citenamefont {Lester}()}]{oxbridge_kinetics}%
  \BibitemOpen
  \bibfield  {author} {\bibinfo {author} {\bibfnamefont {C.}~\bibnamefont
  {Lester}},\ }\href@noop {} {\enquote {\bibinfo {title} {{$M_{T2}$/Stransverse
  Mass/Oxbridge Kinetics Library}},}\ }\bibinfo {howpublished}
  {\url{http://www.hep.phy.cam.ac.uk/~lester/mt2/}},\ \bibinfo {note}
  {{Accessed: January 2016}}\BibitemShut {NoStop}%
\bibitem [{\citenamefont {Cheng}\ and\ \citenamefont
  {Han}(2008)}]{Cheng:2008hk}%
  \BibitemOpen
  \bibfield  {author} {\bibinfo {author} {\bibfnamefont {H.-C.}\ \bibnamefont
  {Cheng}}\ and\ \bibinfo {author} {\bibfnamefont {Z.}~\bibnamefont {Han}},\
  }\href {\doibase 10.1088/1126-6708/2008/12/063} {\bibfield  {journal}
  {\bibinfo  {journal} {JHEP}\ }\textbf {\bibinfo {volume} {12}},\ \bibinfo
  {pages} {063} (\bibinfo {year} {2008})},\ \Eprint
  {http://arxiv.org/abs/0810.5178} {arXiv:0810.5178 [hep-ph]} \BibitemShut
  {NoStop}%
\bibitem [{\citenamefont {{Gu, Jiayin}}()}]{mt2w}%
  \BibitemOpen
  \bibfield  {author} {\bibinfo {author} {\bibnamefont {{Gu, Jiayin}}},\
  }\href@noop {} {\enquote {\bibinfo {title} {{\code{MT2W-1.00a.zip}}},}\
  }\bibinfo {howpublished}
  {\url{https://sites.google.com/a/ucdavis.edu/mass/}},\ \bibinfo {note}
  {{Accessed: January 2016}}\BibitemShut {NoStop}%
\bibitem [{\citenamefont {Fowlie}(2016)}]{api}%
  \BibitemOpen
  \bibfield  {author} {\bibinfo {author} {\bibfnamefont {A.}~\bibnamefont
  {Fowlie}},\ }\href@noop {} {\enquote {\bibinfo {title} {{LHCO\_reader}},}\
  }\bibinfo {howpublished} {\url{http://lhco-reader.readthedocs.org}} (\bibinfo
  {year} {2016}),\ \bibinfo {note} {{Accessed: January 2016}}\BibitemShut
  {NoStop}%
\bibitem [{\citenamefont {Fowlie}\ and\ \citenamefont
  {Marzola}(2015)}]{Fowlie:2014iua}%
  \BibitemOpen
  \bibfield  {author} {\bibinfo {author} {\bibfnamefont {A.}~\bibnamefont
  {Fowlie}}\ and\ \bibinfo {author} {\bibfnamefont {L.}~\bibnamefont
  {Marzola}},\ }\href {\doibase 10.1016/j.nuclphysb.2015.03.025} {\bibfield
  {journal} {\bibinfo  {journal} {Nucl. Phys.}\ }\textbf {\bibinfo {volume}
  {B894}},\ \bibinfo {pages} {588} (\bibinfo {year} {2015})},\ \Eprint
  {http://arxiv.org/abs/1412.5587} {arXiv:1412.5587 [hep-ph]} \BibitemShut
  {NoStop}%
\bibitem [{\citenamefont {Khachatryan}\ \emph {et~al.}(2014)\citenamefont
  {Khachatryan} \emph {et~al.}}]{Khachatryan:2014dka}%
  \BibitemOpen
  \bibfield  {author} {\bibinfo {author} {\bibfnamefont {V.}~\bibnamefont
  {Khachatryan}} \emph {et~al.} (\bibinfo {collaboration} {CMS}),\ }\href
  {\doibase 10.1140/epjc/s10052-014-3149-z} {\bibfield  {journal} {\bibinfo
  {journal} {Eur. Phys. J.}\ }\textbf {\bibinfo {volume} {C74}},\ \bibinfo
  {pages} {3149} (\bibinfo {year} {2014})},\ \Eprint
  {http://arxiv.org/abs/1407.3683} {arXiv:1407.3683 [hep-ex]} \BibitemShut
  {NoStop}%
\end{thebibliography}%
\end{document}